\documentclass[fleqn,usenatbib]{mnras}

\usepackage{mathptmx}

\usepackage[T1]{fontenc}

\DeclareRobustCommand{\VAN}[3]{#2}
\let\VANthebibliography\thebibliography
\def\thebibliography{\DeclareRobustCommand{\VAN}[3]{##3}\VANthebibliography}

\usepackage{graphicx}   
\usepackage{amsmath}    
\usepackage{amssymb}    
\usepackage{subcaption}

\newcommand{\snn}{{\sc SuperNNova }}
\newcommand{\snnns}{{\sc SuperNNova}}
\newcommand{\ssnn}{{\sc SNN }}

\defcitealias{Moller:2022}{M22}
\newcommand{\loose}{\textit{SNN>0.5 }}

\title[DES SNe Ia without redshifts]{The Dark Energy Survey 5-year photometrically classified type Ia supernovae without host-galaxy redshifts}

\author[Möller et al.]{
A. Möller $^{1,2}$\thanks{E-mail: amoller@swin.edu.au},
P.~Wiseman$^{3}$,
M.~Smith$^{4}$,
C.~Lidman$^{5,6}$,
T.~M.~Davis$^{7}$,
R.~Kessler$^{8,9}$,
M.~Sako$^{10}$,
\newauthor
M.~Sullivan$^{3}$,
L.~Galbany$^{11,12}$,
J.~Lee$^{10}$,
R.~C.~Nichol$^{13}$,
B.~O.~S\'anchez$^{14}$,
M.~Vincenzi$^{15}$,
B.~E.~Tucker$^{5}$,
\newauthor
T.~M.~C.~Abbott$^{16}$,
M.~Aguena$^{17}$,
S.~Allam$^{18}$,
O.~Alves$^{19}$,
F.~Andrade-Oliveira$^{19}$,
D.~Bacon$^{20}$,
\newauthor
E.~Bertin$^{21,22}$,
D.~Brooks$^{23}$,
A.~Carnero~Rosell$^{17,24,25}$,
F.~J.~Castander$^{11,12}$,
S.~Desai$^{26}$,
H.~T.~Diehl$^{18}$,
\newauthor
S.~Everett$^{27}$,
I.~Ferrero$^{28}$,
D.~Friedel$^{29}$,
J.~Frieman$^{8,18}$,
J.~Garc\'ia-Bellido$^{30}$,
E.~Gaztanaga$^{11,12,20}$,
\newauthor
G.~Giannini$^{8,31}$,
R.~A.~Gruendl$^{29,32}$,
G.~Gutierrez$^{18}$,
S.~R.~Hinton$^{7}$,
D.~L.~Hollowood$^{33}$,
K.~Honscheid$^{34,35}$,
\newauthor
D.~J.~James$^{36}$,
K.~Kuehn$^{37,38}$,
O.~Lahav$^{23}$,
S.~Lee$^{27}$,
J.~L.~Marshall$^{39}$,
J. Mena-Fern{\'a}ndez$^{40}$,
\newauthor
F.~Menanteau$^{29,32}$,
R.~Miquel$^{31,41}$,
J.~Myles$^{42}$,
R.~L.~C.~Ogando$^{43}$,
A.~Palmese$^{44}$,
A.~Pieres$^{17,43}$,
\newauthor
A.~A.~Plazas~Malag\'on$^{45,46}$,
A.~Roodman$^{45,46}$,
E.~Sanchez$^{47}$,
D.~Sanchez Cid$^{47}$,
I.~Sevilla-Noarbe$^{47}$,
\newauthor
E.~Suchyta$^{48}$,
M.~E.~C.~Swanson$^{29}$,
G.~Tarle$^{19}$,
D.~L.~Tucker$^{18}$,
A.~R.~Walker$^{16}$,
N.~Weaverdyck$^{19,49}$,
\newauthor
L.~N.~da Costa$^{17}$,
M.~E.~S.~Pereira$^{50}$\\
\textit{Affiliations are listed at the end of the paper}
}

\date{Accepted XXX. Received YYY; in original form ZZZ}

\pubyear{2020}

\begin{document}
\label{firstpage}
\pagerange{\pageref{firstpage}--\pageref{lastpage}}
\maketitle

\begin{abstract}

Current and future Type Ia Supernova (SN Ia) surveys will need to adopt new approaches to classifying SNe and obtaining their redshifts without spectra if they wish to reach their full potential. We present here a novel approach that uses only photometry to identify SNe Ia in the 5-year Dark Energy Survey (DES) dataset using the \snn classifier. Our approach, which does not rely on any information from the SN host-galaxy, recovers SNe Ia that might otherwise be lost due to a lack of an identifiable host. We select $2{,}298$ high-quality SNe Ia from the DES 5-year dataset an almost complete sample of detected SNe Ia. More than 700 of these have no spectroscopic host redshift and are potentially new SNIa compared to the DES-SN5YR cosmology analysis. To analyse these SNe Ia, we derive their redshifts and properties using only their light-curves with a modified version of the SALT2 light-curve fitter. Compared to other DES SN Ia samples with spectroscopic redshifts, our new sample has in average higher redshift, bluer and broader light-curves, and fainter host-galaxies. Future surveys such as LSST will also face an additional challenge, the scarcity of spectroscopic resources for follow-up. When applying our novel method to DES data, we reduce the need for follow-up by a factor of four and three for host-galaxy and live SN respectively compared to earlier approaches. Our novel method thus leads to better optimisation of spectroscopic resources for follow-up.

\end{abstract}




\section{Introduction}
Type Ia Supernovae (SNe Ia) are crucial tools to directly measure the cosmic expansion and constrain Dark Energy models. Surveys such as the Dark Energy Survey (DES) and Zwicky Transient Facility (ZTF) have already discovered thousands of SNe Ia and other optical transients \citep{Bellm_2018,Bernstein:2011}. The upcoming Vera C. Rubin Observatory will provide up to 10 million transient and variable detections every night \citep[Rubin, ][]{LSST:2009}. During its 10 year Legacy Survey of Space and Time (LSST) it will detect more than a million SNe which can be used to make precise measurements of the equation-of-state parameter of Dark Energy. To constrain cosmological parameters, SNe Ia first need to be accurately classified and redshifts need to be determined. 

Traditionally, classification of SNe for cosmology is done using real-time spectroscopy as in the DES 3-year analysis and Pantheon+ \citep{DES:3yr,Brout:2022,Scolnic:2018,Betoule:2014,Doi:2010,Contreras:2010,Hicken:2009}. However, spectroscopic resources are limited and thus, a large fraction of detected SNe have not been classified in these datasets. To fully exploit the power of these current and future time-domain surveys, it has become necessary to classify astrophysical objects using photometry instead of the resource-limited spectroscopy. In recent years, many methods have been developed to classify transients using photometry, with an emphasis on supernovae \citep[PSNID, SNLSPC, \snnns, RAPID, SuperRAENN, SCONE;][]{Sako:2011,Moller:2016,Moller:2020,Muthukrishna:2019,Villar:2019,Villar:2020,Qu:2021}.

The DES 5-year cosmology analysis \citep{DES:5yr}, uses photometric instead of spectroscopic classification to obtain the largest high-redshift SNe Ia sample from a single survey \citep{Moller:2022,Vincenzi:2024}. 1499 SNe Ia were classified using their light-curves and spectroscopic host-galaxy redshift information. In
contrast to most previous cosmological samples, SN Ia classification probabilities were incorporated in the cosmology analysis \citep{Moller:2020,Qu:2021,Hlozek:2012,Vincenzi:2022}. This analysis provides the tightest cosmological constraints by any supernova dataset to date. It also overcomes contamination uncertainties from previous photometrically classified cosmology analyses \citep{Jones:2018}. 

To obtain even larger samples and reduce selection biases, methods have been extended to ignore all spectroscopic information. Most of these methods use complete light-curves and either photometric host-galaxy redshifts or photometric SN-derived redshifts \citep{Bazin:2011,Moller:2016,Lochner:2016,Carrick:2021,Boone:2021, Gagliano:2023}. Some of these methods have been used for obtaining cosmological constraints \citep{Chen:2022,Ruhlmann-Kleider:2022}. However, precise classification without the use of any redshift information remains a challenge in particular when using early light-curves \citep{Moller:2021,Leoni:2022,Moller:2022BNN}. 

In this work, we classify SNe Ia using only the information from the 5-year DES light-curves using an extension of the machine learning framework \snn \citep{Moller:2020}. We aim to fully harness the power of the DES data, by identifying most of the detected SNe Ia in this survey, regardless of whether or not a host redshift has been acquired. We exploit the improved statistics that come from larger, more complete, and more representative samples. 

To use these SNe Ia for cosmology, rates, and other astrophysical analyses, we require both accurate classification and redshifts. Traditionally redshifts are obtained from spectra from the SN or host-galaxies using spectroscopic follow-up \citep{Smith:2018, Lidman:2020}. An alternative is to use host-galaxy photometric redshifts but these are biased and have not been widely used in cosmological analyses \citep{Ruhlmann-Kleider:2022}. A promising avenue is to use a subsample of host-galaxies that have highly accurate photometric redshifts such as Luminous Red Galaxies \citep{Chen:2022}. However, for these methods, host-galaxies need to be identified and high SNR photometry acquired or produced with stacked images. An alternative, which does not require host identification, is to infer redshifts from the SN light-curves directly. These methods have been explored with data from previous surveys obtaining promising results \citep{Sako:2011, Palanque-Delabrouille:2010, Kessler:2010FittedPhotoZ}. In this work, we derive redshifts from SN light-curves using the SNphoto-z method \citep{Kessler:2010FittedPhotoZ}, assess biases and the impact these biases have on astrophysical analyses.

Future surveys will continue to detect more SNe than it is possible to follow-up spectroscopically both for classification and host-galaxy redshift acquisition. In the case of Rubin, the 4-metre Multi-Object Spectroscopic Telescope (4MOST) Time-Domain Extragalactic Survey \citep[TiDES;][]{Swann:2019} will aim to classify live SNe and obtain host-galaxy redshifts for cosmology up to a limiting magnitude of $22.5$. 4MOST still won't be able to follow up all SNe and transients from Rubin.

With a focus towards future surveys and their spectroscopic follow-up programmes, here we use DES data as a test bench to explore the optimisation of follow-up resources for both host-galaxy redshift acquisition and live supernovae follow-up. The main spectroscopic follow-up provider for DES was the Australian Dark Energy Survey (OzDES) on the 3.9-m Anglo-Australian Telescope \citep{Yuan:2015,Childress:2017,Lidman:2020}. OzDES targets were prioritised using a template fitting method called Photometric Supernova IDentification software \citep[PSNID]{Sako:2011} and selecting hosts mostly with ${r}<24$. However, this method is time intensive and it will be difficult to scale it for future surveys. To address this, machine learning algorithms have been developed for this challenging task \citep{Muthukrishna:2019,Moller:2020,Leoni:2022}. In this work, we use \snn \citep{Moller:2020}, a photometric classification framework, for spectroscopic follow-up optimisation using DES data.

This paper is organised as follows. We introduce the Dark Energy Survey in Section~\ref{sec:DES}. For light-curve classification, we use the algorithm \snn introduced in Section~\ref{sec:snn_noz}. This algorithm is trained on realistic DES simulations on both complete and partial light-curves with performances on complete and partial shown in Sections~\ref{sec:sims_complete} and ~\ref{sec:sims_early}, respectively. In Section~\ref{sec:SALT}, we use the simulations described in Section~\ref{sec:snn_noz} to examine the SNphoto-z estimation and its biases which will be used for sample analysis but not for classification. In Section~\ref{sec:HQsample} we select a SN Ia sample without the use of any redshift information, study its properties and compare it to previous DES SN Ia samples. We then explore how machine learning classification can improve follow-up optimisation for host-galaxies in Section~\ref{sec:fuphost} and for early SN identification using partial light-curves in Section~\ref{sec:early}. We conclude with prospects for future surveys such as Rubin LSST and 4MOST in Section~\ref{sec:rubin}.

\begin{figure*}
\centering
    \includegraphics[width=\textwidth]{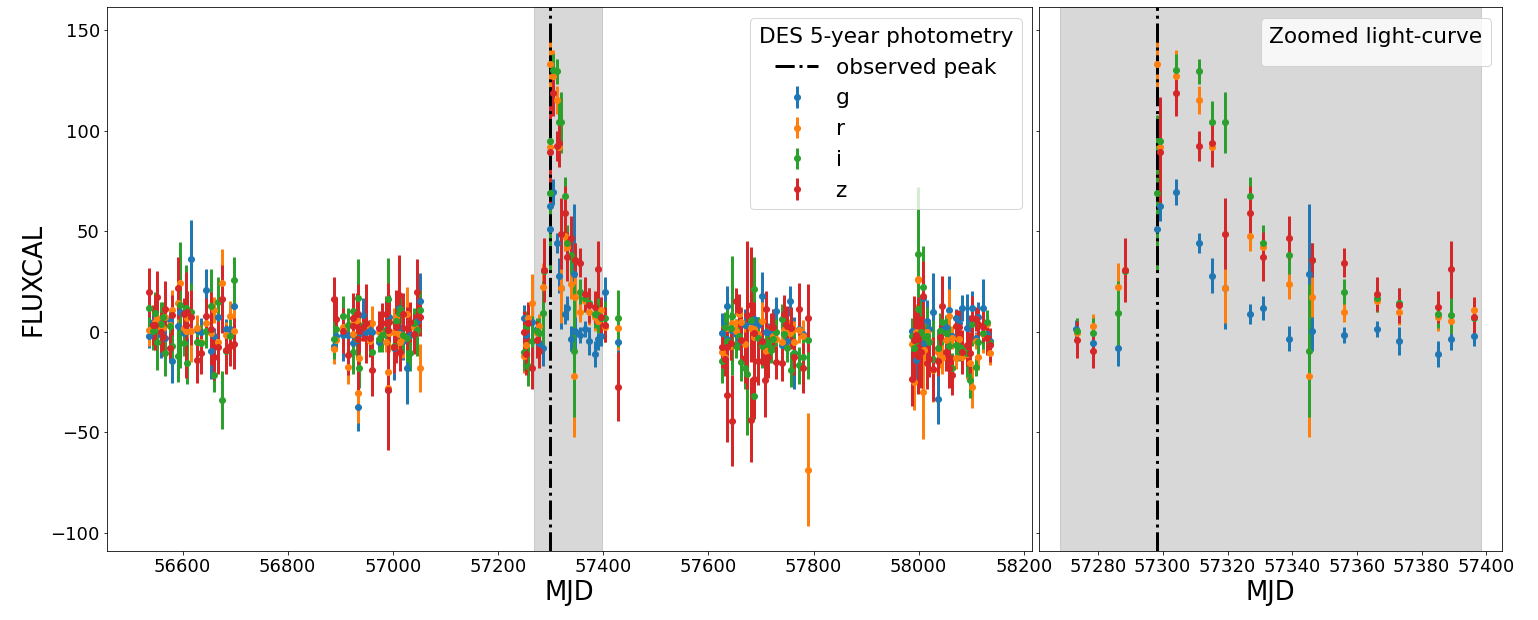}
    \caption{Light-curve of DES15X2kvt. The measured calibrated flux (FLUXCAL, defined in Section~\ref{sec:sims}) in $g, r, i$ and  $z$ bands is plotted against Modified Julien Date (MJD). In the left panel we show the full 5-year light-curve. In the right panel we show the light-curve 30 days before to 100 days after the observed peak flux.}
    \label{fig:lc_peak}
\end{figure*}

\section{Dark Energy Survey (DES)}\label{sec:DES}

In this work, we select SNe Ia using only light-curve information from the Dark Energy Survey. DES was a photometric survey that used the Dark Energy Camera \citep[DECam;][]{Flaugher:2015} at the Victor M. Blanco Telescope in Chile. It consisted of a wide-area survey (DES-wide) and a supernova survey (DES-SN). DES-SN, which is used in this work, imaged ten $2.7~{\rm deg}^2$ fields with an average cadence of $7$ days in the $griz$ filters during 5 years \citep{Abbott:2018}. Eight of these ten fields (X1, X2, E1, E2, C1, C2, S1, and S2) were observed to a single-visit depth of $m\approx 23.5$ mag (\lq shallow fields\rq), and the other two (X3,C3) were observed to a depth of $m\approx24.5$ mag (\lq deep fields\rq). Detailed information on the SN survey can be found in \cite{Smith:2020}.

Transients were identified using the DES Difference Imaging Pipeline {\sc diffimg} \citep{Kessler:2015} coupled with a machine learning algorithm \citep{Goldstein:2015} to reduce difference imaging artefacts. A candidate SN was defined from the difference images by requiring at least two detections with effective S/N threshold about 5 in each band. These criteria were designed to remove artefacts and asteroids. This yielded a sample containing $31,636$ light-curves with 5-year photometry. An example of a light-curve is shown in Figure~\ref{fig:lc_peak}.

From this DES SN candidate sample, SNe Ia were selected for the DES 5-year cosmological analysis \citep{DES:5yr}. Instead of spectroscopic selection \citep{Smith:2020}, SNe Ia were weighted by their probability of being SNe Ia from the classification framework \snn \citep{Moller:2020} using light-curves and host-galaxy spectroscopic redshifts \citep[][hereafter M22]{Moller:2022}. This SNe Ia sample is the largest and deepest SN cosmological sample acquired from a single survey. Photometric misclassification was shown not to be a limiting uncertainty in the cosmological analysis \citep{Vincenzi:2022, Vincenzi:2024}. Part of this analysis tested other photometric classifiers such as SCONE \citep{Qu:2021} to evaluate the systematic uncertainty.

A subsample of DES SNe Ia were classified using spectroscopic follow-up. For this, potential SNe were identified early (before or around maximum brightness). A trigger is defined as a sequence of detections that results in tracking the light-curve with forced-photometry
and consideration for  spectroscopic follow-up. SDSS required 2 detections on 2 separate nights;
DES required 1 (or more) detections on 2 separate nights \citep{Sako:2011}, and Rubin LSST
will require just 1 detection. In Section~\ref{sec:early} we explore early classification with different triggers.

In this work, we use the DES SN candidate sample to select SNe Ia without any spectroscopic information from either host or the SN. We only use the SN candidates 5-year photometric light-curves.

\section{Classification performance on simulations} \label{sec:snn_noz}
We make use of \snn ({\sc SNN}) to select SN Ia candidates \citep{Moller:2020}. \ssnn is an open-source light-curve classification framework which was used for the classification of Type Ia SNe in the DES 5-year cosmological analysis using light-curves and host-galaxy redshifts \citep{DES:5yr, Moller:2022} and is part of the Rubin broker {\sc Fink} \citep{Moller:2021, Fraga:2024}. 

\ssnn is a non-parametric method that uses as input fluxes and their measurement uncertainties over time for light-curve classification. Additional information such as host-galaxy redshifts can be included to improve performance such as in \cite{DES:5yr}. \ssnn  includes different classification algorithms, such as LSTM\footnote{Long short-term memory \citep[LSTM;][]{Hochreiter:1997}} Recurrent Neural Networks (RNNs) and two approximations for Bayesian Neural Networks.  These algorithms can be trained for binary or multi-class classification and then applied to independent datasets to obtain probabilities of a light-curve being of a certain class. The classification probabilities can be used to select a sample by performing a threshold cut or by weighting the contribution of candidates by their classification score as in the BEAMS and BBC methods \citep{DES:5yr,Vincenzi:2024,Moller:2022}. In this work we use a SN Ia probability threshold which we will denote as \textit{SNN>threshold}.

In this work we train \ssnn for classification of SNe Ia vs non Ia using only photometric measurements. To avoid luminosity biases, we use the \textit{cosmo\_quantile} normalisation as in \citetalias{Moller:2022} which, for a given light curve, normalises fluxes and uncertainties by the 99th quantile of the flux distribution (to avoid using an outlier). This normalizes the fluxes for each light curve to 1 or near 1, thus making the classification model agnostic to the relative differences in apparent brightness between SNe and retains colour and signal-to-noise information for the classification. A thorough study of the cosmological biases from \ssnn classification can be found in \cite{Vincenzi:2022}.

The classifier was trained using DES-like simulations described in Section~\ref{sec:sims} and the \snn configuration in \citetalias{Moller:2022}. The performance obtained for complete light-curves (using all SN photometry) is discussed in Section~\ref{sec:sims_complete} and for partial light-curves (using photometry before maximum brightness) in Section~\ref{sec:sims_early}.

\subsection{Simulations} \label{sec:sims}

DES-like simulations are used to train and test our photometric classifier using only light-curves. Simulations contain light-curves of different SNe types generated with realistic observing conditions. These simulations also include a host redshift, however we withhold this information from the \snn classifier. Details on the simulations, which were generated using SNANA \citep{Kessler:2009} within the PIPPIN framework \citep{Hinton:2020}, can be found in \citetalias{Moller:2022} and \cite{Kessler:2019}. Throughout this work, we use SNANA calibrated flux (FLUXCAL) defined from the magnitude with a fixed zeropoint given by: $mag = 27.5 - 2.5*log_{10}$(FLUXCAL).

As in \citetalias{Moller:2022} we first create a training sample with the same number of Type Ia and core-collapse SNe after trigger and selection requirements (equivalent to 50\% type Ia and 50\% core-collapse SNe). This balanced training sample contains $3.6\times10^6$ SNe and covers the redshift range from $0.05$ to $1.3$. As in \cite{Vincenzi:2022}, it contains Type Ia based on models in \cite{Guy:2007} and the optical+NIR extension from \cite{Pierel:2018}, peculiar Ia \citep[SN1991bg- like SNe and SN2002cx-like SNe;][]{Kessler:2019simsplasticc} and core-collapse SNe from \cite{Vincenzi:2019} using volumetric rates from \cite{Frohmaier:2019}.

We generate a smaller data-sized simulation to estimate the expected number of SNe Ia in the DES survey as well to test our photometric classifiers. We simulate 30 realisations of the DES survey using the expected rates of type Ia and non Ia SNe. This simulation contains $\approx 60\%$ type Ia and $40\%$ core-collapse SNe, and was generated using the expected abundances of different types of supernovae through cosmic time. 

\subsection{Performance on complete light-curves} \label{sec:sims_complete}

We evaluate the classification of complete light-curves: up to hundred days beyond
the time of peak brightness. We use accuracy, efficiency and purity as metrics to assess the performance of the classifier.

Accuracy is measured as the number of correct predictions against the total number of predictions. More explicitly, it is calculated as follows:
  \begin{equation}
  \mathrm{accuracy} = \frac{\mathrm{TP+TN}}{\mathrm{TP+TN+FP+FN}}
  \end{equation}
where TP (resp. TN) are true positives (resp. true negatives) and FP (resp. FN) are false positives (resp. false negatives). TP is the number of correctly classified SNe Ia while TN is the number of correctly classified non SNe Ia.

The purity of the SN Ia sample and the classification efficiency are defined as:
  \begin{equation}
  \mathrm{purity} = \frac{\mathrm{TP}}{\mathrm{TP+FP}}; \quad \mathrm{efficiency} = \frac{\mathrm{TP}}{\mathrm{TP+FN}}
  \end{equation}\label{eq:purity}

In Table~\ref{tab:snnsims_nz} we list the accuracies, purities, and efficiencies obtained for the balanced dataset (same number of Type Ia and core-collapse SNe) and the more realistic DES test set. The balanced dataset is useful as an evaluation of the machine learning algorithm while the test dataset can be used to assess the reliability of the selected sample as it is physically more representative. We find high-accuracies, purities and efficiencies for both datasets.


As in \citetalias{Moller:2022}, we use ensemble predictions to select our sample. In Table~\ref{tab:snnsims_nz}, we obtain predictions with different \snn models trained with different initiation parameters (random seed) and average them to obtain an "ensemble probability". Here we use 5 models, also called an "ensemble set", trained with different seeds. To report the performance of the methods, we quote the mean and standard deviation of a given metric using 3 ensemble sets.

\begin{table}
    \caption{Type Ia vs. non Ia classification metrics for complete light-curves with no redshift information. The model was trained and evaluated using two datasets: balanced and test. The metrics indicate the performance of the ML classifier. The metrics for the test dataset indicate the expected performance in a real survey. We show the single model and the ensemble method metrics. Uncertainties for the single model are computed from the variance of 5 models with different seeds and uncertainties for the ensemble methods are computed using three ensembles of fives seeds. }
    \label{tab:snnsims_nz}
\begin{tabular}{llll}
              method &         accuracy &       efficiency &           purity \\
\hline
\multicolumn {4} {c}{ \textit{balanced dataset}}\\
\hline
              single model &    $97.15\pm 0.03$ &    $97.94\pm 0.06$ &  $96.42\pm 0.07$ \\
      ensemble &    $\mathbf{97.34\pm 0.01}$ &    $\mathbf{98.17\pm 0.02}$ &  $\mathbf{96.57\pm 0.01}$ \\
\hline
\multicolumn {4} {c}{ \textit{test dataset (realistic rates)}}\\
\hline
       single model &      $97.04 \pm 0.02$ & $98.12 \pm 0.06$ &    $97.20 \pm 0.05$ \\
ensemble &      $\mathbf{97.22 \pm 0.01}$ & $\mathbf{98.36 \pm 0.01}$ & $\mathbf{97.30 \pm 0.01}$ \\
\end{tabular}
\end{table}

\begin{table}
    \caption{Type Ia vs. non Ia classification metrics for partial light-curves with no redshift information. These light-curves contain only photometric measurements up to their peak brightness.}
    \label{tab:snnsims_nz_early}
\begin{tabular}{llll}
              method &         accuracy &       efficiency &           purity   \\
\hline
\multicolumn {4} {c}{ \textit{balanced dataset}}\\
\hline
              single model  & $90.4 \pm 0.1$ & $91.5 \pm 0.2$ & $89.4 \pm 0.2$\\
     ensemble &   $90.73 \pm 0.01$ &  $91.9 \pm 0.1$ &  $89.7 \pm 0.1$ \\
\hline
\multicolumn {4} {c}{ \textit{test dataset (realistic rates)}}\\
\hline
       single model &   $90.6 \pm 0.1$ & $92.1 \pm 0.2$ & $91.7 \pm 0.2$ \\
ensemble &      $90.46 \pm 0.03$ & $92.49 \pm 0.03$ & $91.93 \pm 0.03$ \\
\end{tabular}
\end{table}

\subsection{Performance for partial light-curves} \label{sec:sims_early}
We now evaluate the performance of our trained classifier when using simulated partial light-curves. When training {\sc SuperNNova}, we crop light-curves to random time-ranges in the dataset, this produces a classification model robust for both complete and partial light-curve classification.

We evaluate the performance on light-curves that were cropped to only contain photometric measurements until peak brightness in Table~\ref{tab:snnsims_nz_early}. As we use fewer photometric measurements per event, the performance is poorer. However this type of classification can be used for scheduling spectroscopic follow-up before SNe fade away.

In the following, we use the single model classifier as the performance gain for the ensemble classifier is small and current early classification mechanisms use a single model. However, the extension to ensembles can provide a gain if resources are available to deploy multiple models as they are not very computationally expensive.

\begin{figure*}
    \centering
    \includegraphics[width=\textwidth]{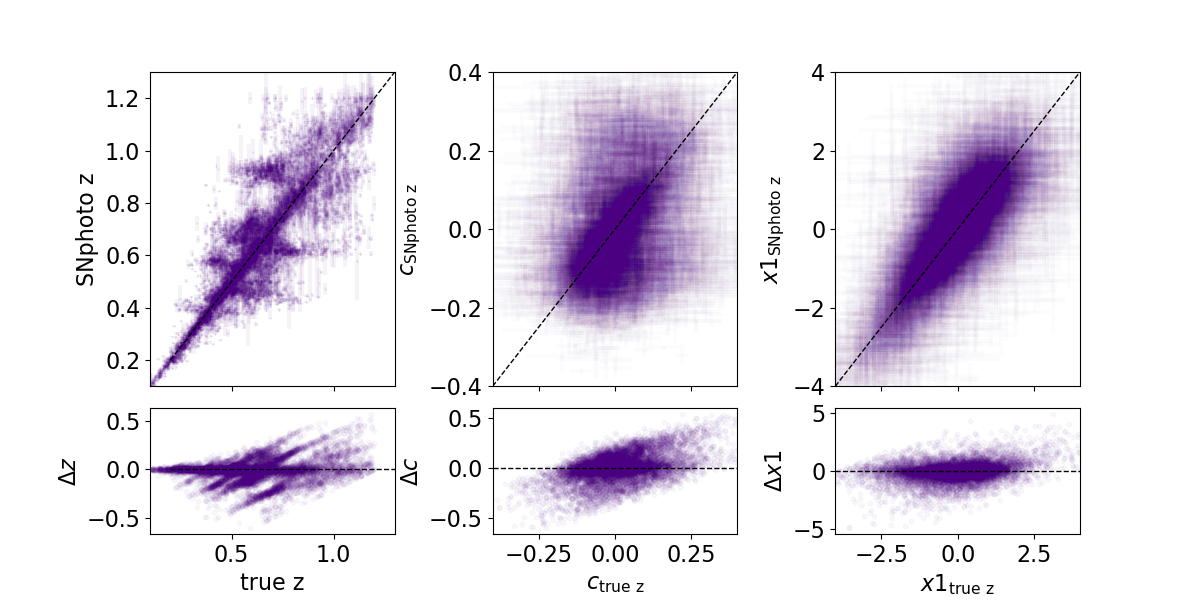}
    \caption{A comparison between the fitted and true parameter values for  simulated SNe Ia. Left: a comparison of the SNphoto-z versus true redshifts. Center and right: comparisons between light-curve parameters colour ($c$) and stretch ($x1$). The dashed line shows the diagonal where the values should lie if they were equivalent. While the simultaneous fits are only slightly biased on average, there is considerable structure, especially in redshift and colour.}
    \label{fig:delta_retro}
\end{figure*}

\section{Estimating redshifts and light-curve parameters simultaneously}\label{sec:SALT}

In this work we will select a photometric SN Ia sample from DES data without the use of redshift information. After classification, we will determine the redshifts and SALT2 light-curve parameters simultaneously on light-curves using the SNphoto-z code described in \cite{Kessler:2010FittedPhotoZ}. 

In this Section, using simulations, we examine biases arising from this fit and evaluate how these biases affect the efficacy of sample cuts in improving the classification efficiency and limiting contamination.

We start by assuming that all the photometrically classified SNe are SNe Ia and fit them with the SALT2 supernova light-curve model based on \citep{Guy:2007} and extended to the optical+NIR \citep{Pierel:2018}. We use the SNANA light-curve fitting program \citep{Kessler:2009} to simultaneously
fit for $z$, $t_0$, $x1$, $c$ and $x_0$; respectively redshift, time of maximum brightness, stretch, colour and amplitude as described in \cite{Kessler:2010FittedPhotoZ}. To obtain better estimates of redshifts for SNe Ia, a weak distance-modulus prior is applied (Appendix~\ref{appendix:distancemusalt})
assuming a $\Lambda$CDM cosmology and we use when available inferred photometric redshifts of the host-galaxies as a Gaussian prior. When no photometric redshift is available, we use a flat prior. We highlight that this SNphoto-z fit uses a cosmological model. 

Detailed analysis of biases on the light-curve parameters and redshift is presented in Section~\ref{sec:simsalt_bias} and their effect on the cuts to improve the classification by limiting contamination in Section~\ref{sec:simsalt_contamination}.

None of the derived redshifts (SNphoto-z) or SALT2 parameters are used for photometric classification. They are only used in Section~\ref{sec:PC_noz_properties} to study the sample properties after classification is done without this information.

\begin{figure}
    \centering
     \includegraphics[width=\columnwidth]{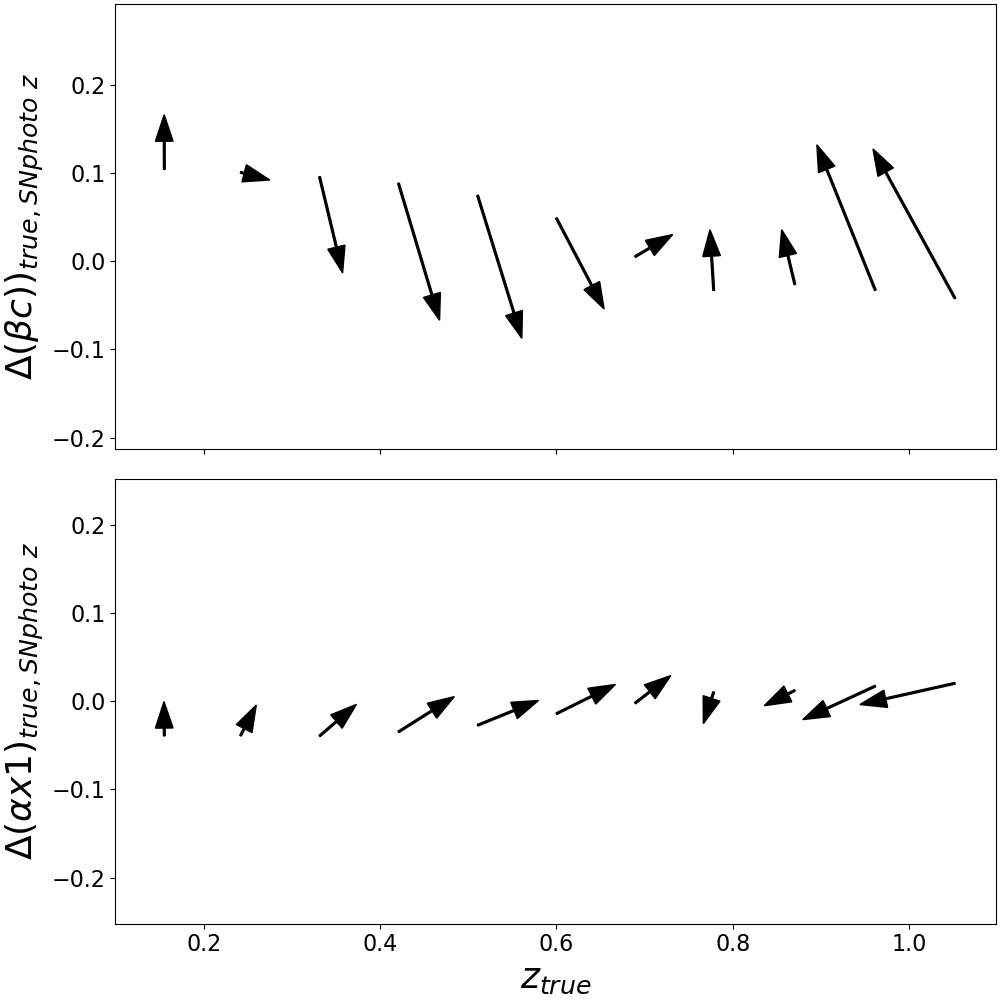}
    \caption{Average offset in SALT2 colour and stretch as a function of redshift. Each arrow represents the average offset in both colour (stretch in the lower plot) and redshift between a fit that fits redshift and SALT light-curve parameters simultaneously and a SALT2 fit that uses as input the true redshift. We show these offsets in magnitudes space as $\beta c$ and $\alpha x1$ where $\alpha = 0.144$ and $\beta = 3.1$. High redshift events are fitted towards lower redshifts, redder colours and lower stretch while intermediate redshift events are offset to higher redshifts, bluer colours and higher stretch.}
    \label{fig:migration_cx1_zHD}
\end{figure}

\subsection{SNphoto-z and light-curve parameters biases} \label{sec:simsalt_bias}

We use the test simulations to evaluate the fitted light-curve parameters and SNphoto-z. In Figure~\ref{fig:delta_retro} we compare the fitted light-curve parameters and SNphoto-z against their true values. 

The fitted parameters are slightly biased on average, with median shift of $-0.003^{+0.035}_{-0.065}$, $0.008^{+0.086}_{-0.058}$ and $-0.0042^{0.315}_{-0.42}$ for redshift, colour and stretch respectively (errors are indicated by the 25th and 75th quantiles). For redshift, colour and stretch, we compute an outlier fraction of 0.1, 0.06 and 0.07 using the interquartile range (IQR) method.

A complex structure can be found in particular for the redshift estimation. \cite{Chen:2022} finds a similar structure, in particular for redshifts around 0.4, when comparing galaxy photometric redshifts obtained in redMaGiC galaxies and their spectroscopic ones. These luminous red galaxies are expected to have highly accurate photometric redshifts and were shown to provide constraints with equivalent Hubble scatter that when using spectroscopic redshifts \citep{Chen:2022}.

In Figure~\ref{fig:migration_cx1_zHD}, we plot the average behaviour of the SNphoto-z and colour/stretch for simulated SNe Ia. We find a pattern of offsets resulting from degeneracies between colour/stretch and redshift. Interestingly, around redshift 0.7 where noise starts dominating the $r$ because the rest-frame UV regions has low flux, only $i,z$ are sampling the light-curve and the offset reverses. Similarly, at redshift around 0.9 the noise dominates the $i$ band thus light-curves are only well sampled in the $z$ band. These effective band drop-outs due to low rest-frame UV flux highlight the importance of multi-band light-curves. These shifts introduce structured systematics. If these simultaneous fits are to be used in further analyses these offsets must be taken into account potentially by the use of bias corrections, a hierarchical model or grouping events in less bias affected bins. An alternative is to use stronger priors for redshift using host-galaxy photometric redshifts to reduce biases. A detailed study in DES of the cosmological biases using photometric redshifts, including SNphoto-z with or without priors, can be found in \cite{Chen:2024}.

\begin{figure}
    \centering
    \includegraphics[width=\columnwidth]{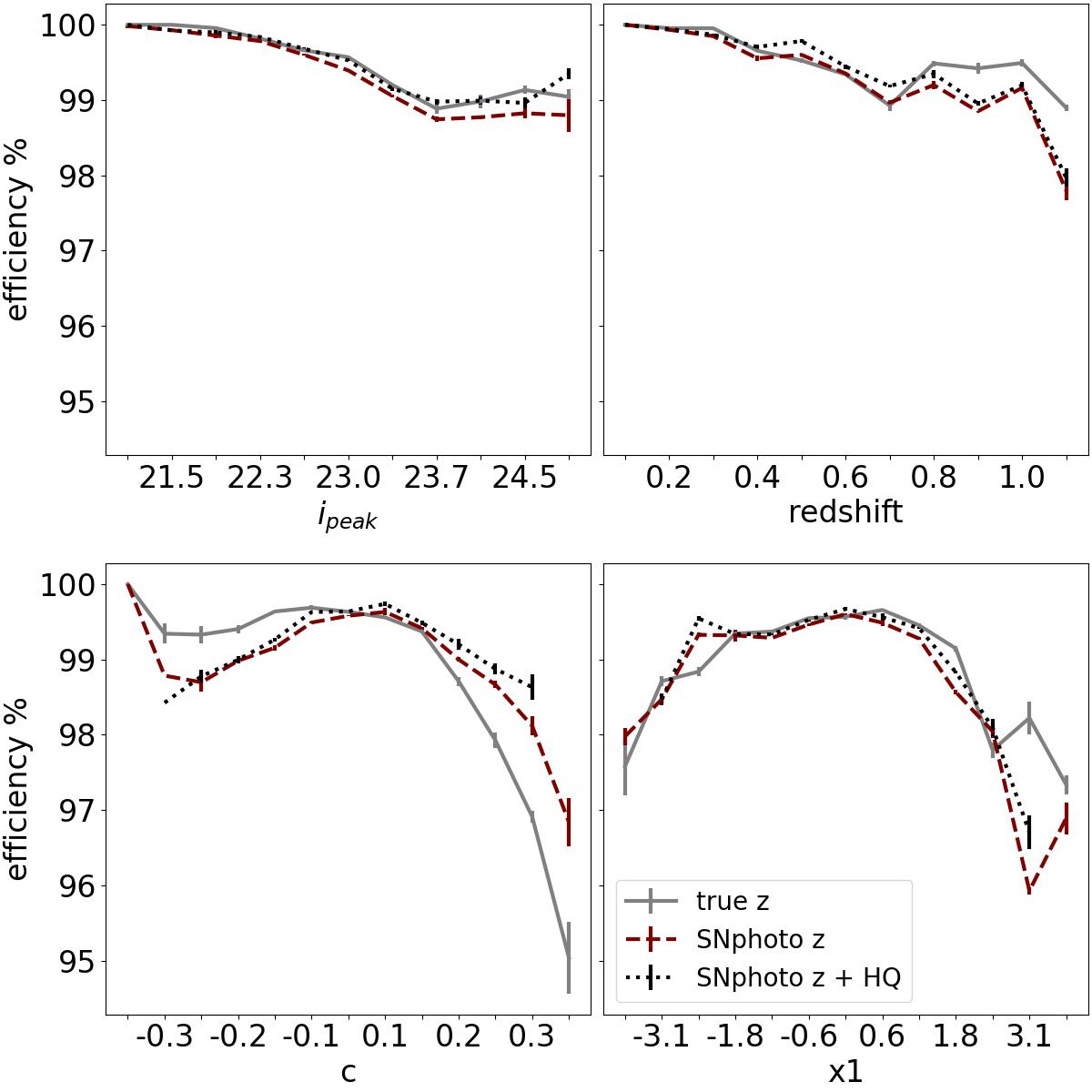}
    \caption{The percentage efficiency per bin of fitted SALT2 peak i-band magnitude ($i_{\rm peak}$), redshift, colour ($c$) and stretch ($x1$) for the simulated dataset. We show the efficiency for a SALT2 fit with a fixed true redshift and that obtained using the SNphoto-z obtained simultaneously with SALT2 light-curve parameters. We also study a sample selected with additional HQ cuts.}
    \label{fig:eff_noz}
\end{figure}

\subsection{The effect of SNphoto-z fit on SNe Ia samples} \label{sec:simsalt_contamination}

In this Section we study how cuts on light-curve parameters affect efficiency and contamination. We study two cuts: the baseline sample selected using only light-curves with a threshold of \loose using the model in Section~\ref{sec:sims_complete}, and a high-quality (HQ) sample with additional cuts on the light-curve parameters. The latter aims to mimic samples for cosmology that apply extra cuts to reduce peculiar SNe Ia \citep{Vincenzi:2020}. The HQ cuts are: $-3.0<x1<3.0$, $-0.3<c<0.3$, and $\sigma_{x1}<1$ and $\sigma_{t_0}<2$. Where $c$, $x1$, $\sigma_{t_0}$ are estimated using SALT2 light-curve fit and represent colour, stretch and the error on the time of maximum light respectively. We also require that the SALT2 chi-square fit probability is larger than $0.001$ cut as in \citetalias{Moller:2022}.

In Figure~\ref{fig:eff_noz} we show the true and measured efficiency as defined in Equation~\ref{eq:purity} for 3 cases: the \loose sample using its true redshift, the \loose sample using SNphoto-z, and a HQ sample using SNphoto-z. In general, we find classification efficiencies above 98\% for most of the parameter space. The samples show higher measured efficiency for SNphoto-z  due to the migration of true bluer events to redder ones. Conversely, the measured efficiency is lower at higher redshifts.

\begin{figure}
    \centering
    \includegraphics[width=\columnwidth]{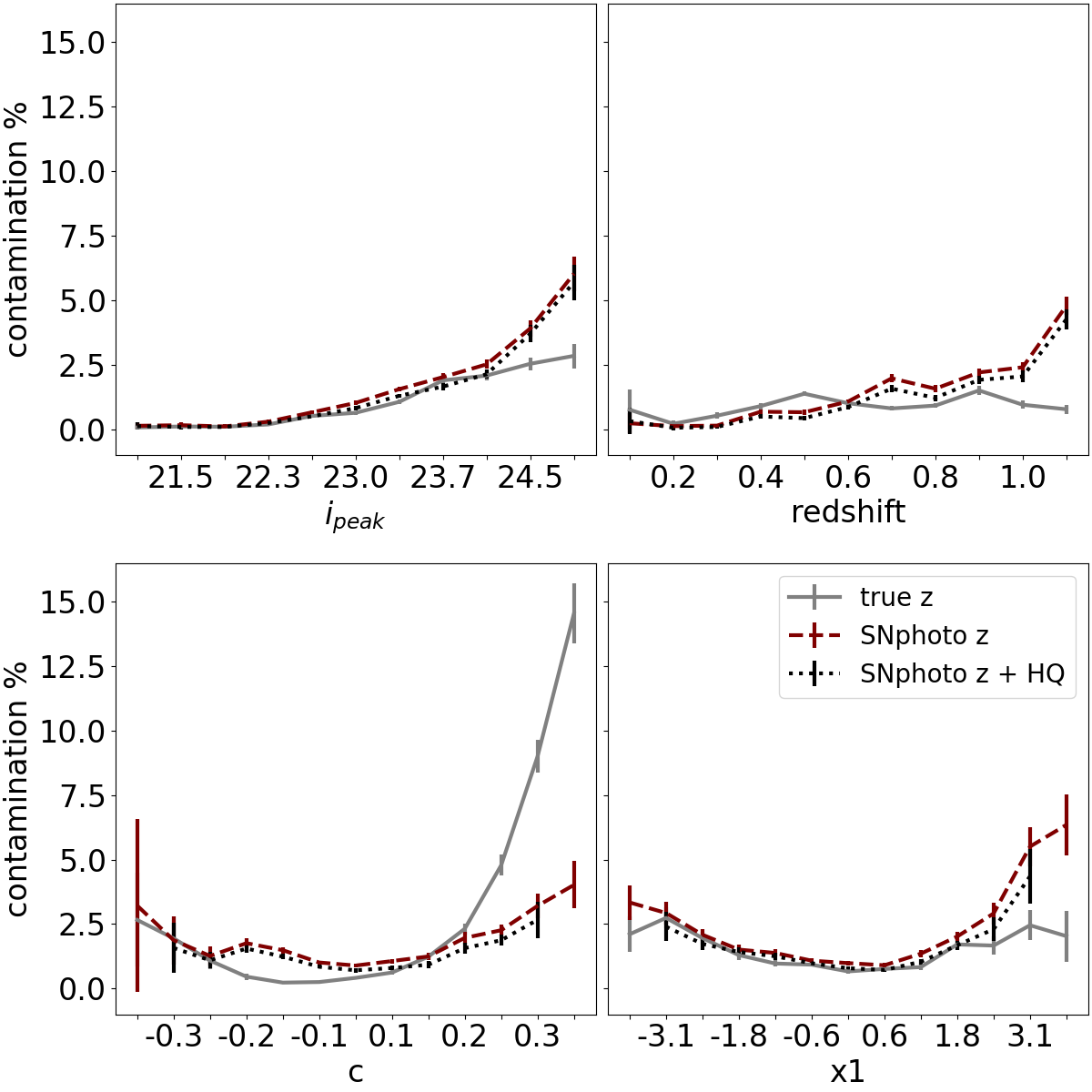}
    \caption{The percentage contamination per bin of fitted SALT2 peak i-band magnitude ($i_{\rm peak}$), redshift, colour ($c$) and stretch ($x1$) for the simulated dataset. We show the contamination for a SALT2 fit with a fixed true redshift and that obtained with SNphoto-z fitted simultaneously with SALT2 parameters. We also study a sample selected with the simultaneous fit and HQ cuts. Contamination is distributed differently when using a fixed or SNphoto-z.}
    \label{fig:cont_noz}
\end{figure}

We also study contamination as a function of light-curve properties in Figure~\ref{fig:cont_noz}. Contamination is measured $1-purity$ as defined in Equation~\ref{eq:purity}. The overall contamination is less than $6\%$ in any parameter bin while the true contamination is higher for redder events. However, when measuring it using the SNphoto-z, this contamination migrates to other colour bins and can also be absorbed by the lack of convergence of the fit. Higher contamination for redder events has also been observed for samples selected using host-galaxy redshifts such as in \cite{Vincenzi:2022, Moller:2022}. When using SNphoto-z, we find that more distant and hence fainter supernovae have a higher contamination.

For the purpose of using this sample for astrophysical analyses, it is promising that the contamination of a sample using SNphoto-zs remains low and below $6\%$ for any given parameter bin. Applying HQ cuts reduces this contamination for the complete parameter space. This causes only a small reduction in efficiency for higher stretch events and events at higher redshifts.

\begin{table*}
    \caption{The number of candidates selected after each cut is applied. We show results for the shallow and deep fields, as well as the total number. Note that a couple of percent events belong to both shallow and deep fields due to field overlap. Columns show the selection cut, the number of selected candidates, the number of spectroscopic SN Ia in the sample and the number of the DES 5-year photometrically classified SNe Ia (photo Ia M22).}
    \label{tab:PC_noz}
\begin{tabular}{lrrrrrrr}            
Selection cut & \multicolumn{2}{c}{shallow} & \multicolumn{2}{c}{deep} &  \multicolumn{3}{c}{total DES 5-year} \\
 & selected & spec Ia & selected & spec Ia & selected & spec Ia & photo Ia \citetalias{Moller:2022} \\       
\hline 
\hline

DES-SN 5-year candidate sample &              23795 &              322 &           7863 &            93 &           31636 &            415 &                1484 \\
Filtering multi-season &       9607 &              317 &           4464 &            88 &           14070 &            405 &                1484 \\
Photometric sampling &             8969 &              314 &           4150 &            86 &           13118 &            400 &                1484 \\
SNN>0.001 &             3680 &              303 &           1996 &            83 &            5676 &            386 &                1481 \\
SNN>0.5 (high purity) &          2199 &              291 &           1348 &            77 &            3547 &            368 &                1376 \\
Converging SALT2 and SNphoto-z fit &        1630 &              250 &            909 &            60 &            2539 &            310 &                1261 \\
HQ &       1559 &              249 &            739 &            60 &            2298 &            309 &                1236 \\
\end{tabular}
\end{table*}

\section{Classification of SNe Ia without redshift information}\label{sec:HQsample}

In this Section we classify light-curves without redshift information to obtain a large, high quality sample of photometrically selected SNe Ia. First, we use simulations to estimate the expected number of SNe Ia in DES in Section~\ref{sec:estimatedSNeIa}. Next, we pre-process DES data in Section~\ref{sec:preprocessing}. We define a sample using a threshold similar to \citetalias{Moller:2022} in Section~\ref{sec:snn05}. Using the SNphoto-z method introduced in Section~\ref{sec:SALT}, we obtain a high-quality sample in Section~\ref{sec:hq} and study its properties in Section~\ref{sec:PC_noz_properties}. We conclude by comparing this sample to other DES SN Ia samples in Section~\ref{sec:allDESsamples}.

\subsection{Expected number of HQ DES SNe Ia}\label{sec:estimatedSNeIa}
We use the DES realistic simulation introduced in Section~\ref{sec:sims} to estimate the number of SNe Ia the DES survey. This simulation consists of 30 realistic simulations of the full DES 5-year SN survey up to redshift $1.2$.

From these simulations, we expect to detect $4{,}961\pm69$ SNe Ia (median and standard deviation of 30 realisations). No selection cuts other than detection are applied at this stage.

From these, we expect $2{,}360\pm43$ high-quality SNe Ia using the cuts introduced in Section~\ref{sec:simsalt_contamination}. For this estimate, we use the simulated redshift when fitting the light-curve with SALT2. We then apply the cuts.

This simulation also includes other types that are not normal type Ia SNe with realistic rates. We estimate that DES detected $3{,}466 \pm 64$ SNe of other types. Importantly, we estimate up to $231\pm 13$ non normal type Ia SNe that would pass the HQ cuts if a SALT2 fit using their redshift was done. These SNe are contaminants for cosmology analyses which are reduced by using photometric classifiers. For a thorough discussion on biases, refer to \cite{Vincenzi:2022, Vincenzi:2024}.

\subsection{SN candidates pre-processing} \label{sec:preprocessing}

In this Section we use the DES SN candidate sample introduced in Section~\ref{sec:DES}. We make use of light-curves from 31,636 candidates, using both the fluxes and their uncertainties.

We use the pre-processing introduced by \citetalias{Moller:2022} to prepare light-curves for photometric classification with {\sc SuperNNova}:
\begin{itemize}
    \item We select a subset of 5-year photometry within a time-window in the observer frame of 30 days before to 100 days after maximum brightness of the detected event, as shown in Figure~\ref{fig:lc_peak}. 
    \item We eliminate photometry that has been flagged as flawed using bitmap flags from {\sc Source Extractor} \citep{Bertin:1996}. $6\%$ of measurements are discarded here.
    \item We filter multi-season events, which include AGN, by requiring a large ratio of good detections with respect to all detections using a Real/Bogus classifier. We require a ratio between the number of epochs with detections that pass the Real/Bogus classifier \citep[{\sc autoScan;}][]{Goldstein:2015} and the total number of epochs with detections to be larger than 0.2 as in \cite{Smith:2020}.
\end{itemize}

With these cuts, the sample is reduced from $31{,}636$ to $14{,}070$ SN candidates. While these cuts reduce the contamination, some residual AGN remain. The number of candidates that remain after each cut is listed in Table~\ref{tab:PC_noz}. We highlight that from the original 415 spectroscopic SNe Ia, 10 are eliminated due to the multi-season cut as they may be in galaxies with AGN.

Additionally, we require at least one photometric detection before 10 days after peak, and at least one after 10 days after peak. We highlight that the peak brightness is an observed peak brightness and it does not necessarily correspond to the peak SN flux. 5 events do not pass these criteria.

This sample of $13{,}118$ candidates,includes the following spectroscopically classified events: 400 SNe Ia (241 of these were in the DES 3 year analysis), 83 core-collapse SNe, 2 peculiar SNe Ia, 16 Super Luminous SNe, 1 Tidal Disruption Event, 1 M Star and 36 AGN.

\begin{figure*}
    \centering
    \includegraphics[width=\textwidth]{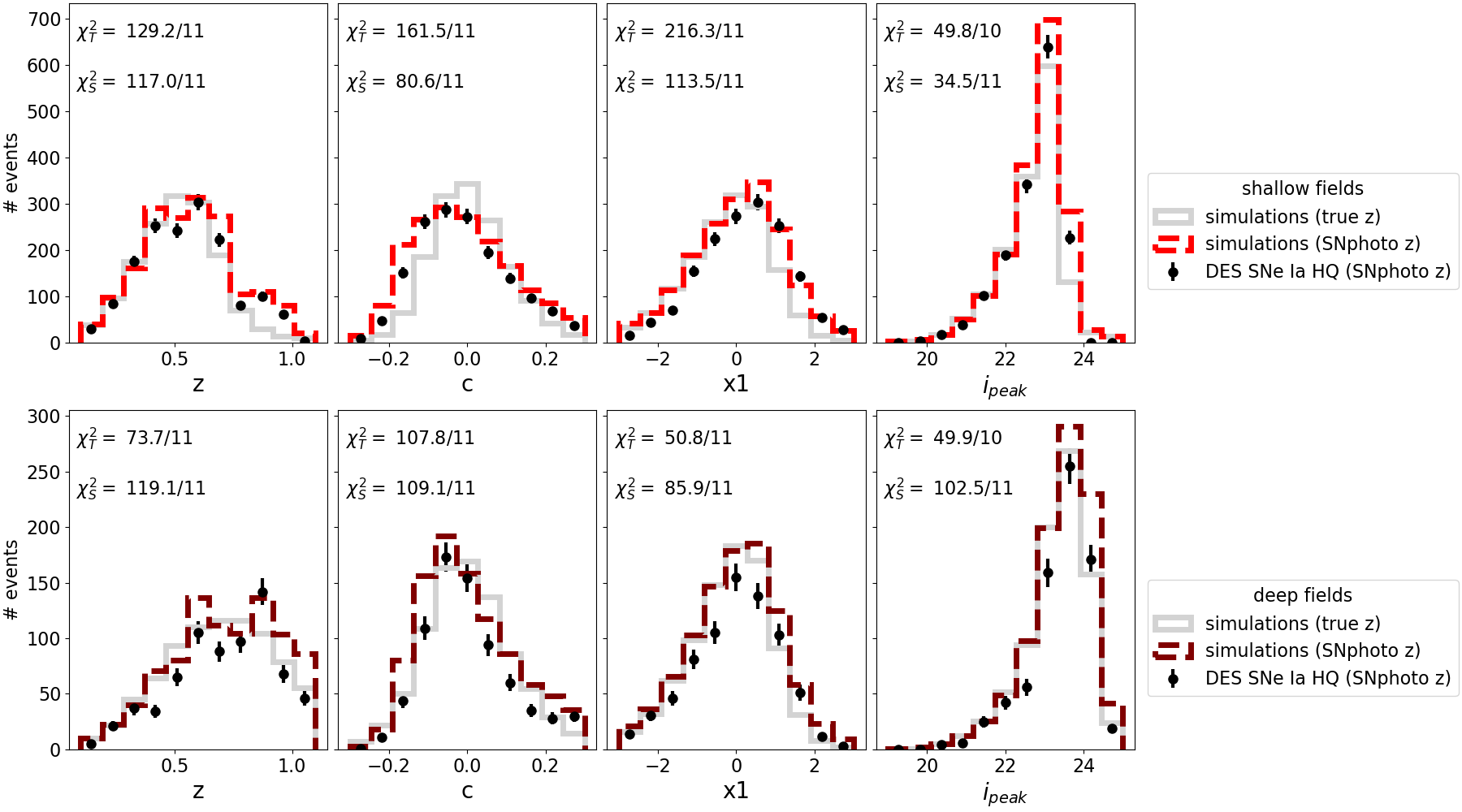}
     \caption{Distributions of redshift, SALT2 $x1$, SALT2 $c$ and i-band peak magnitude $i_{\rm peak}$ for the high quality, photometrically classified sample. We show distributions for both the SNphoto-z and SALT2 light-curve parameters (in red for shallow fields and maroon for deep fields) and the distributions if simulated redshift was used (in grey). Poisson uncertainties are assumed. Both the simulation and data pass HQ cuts. The goodness-of-fit for each histogram is shown as the $\chi^2$/number of bins on each plot for both the SNphoto-z ($\chi^2_S$) and fixed true redshift ($\chi^2_T$). The simulations replicate the data better when the SNphoto-z is used for the shallow fields only. }
    \label{fig:hists_noz}
\end{figure*}

\subsection{High purity sample (SNN>0.5)} \label{sec:snn05}

We select a higher purity sample with the same threshold as \citetalias{Moller:2022} but without the use of redshift information. We select $3{,}545$\footnote{Two light-curves are discarded since they have close-by AGN as discussed in Section~\ref{sec:hq}.} light-curves that have an ensemble probability of being SNe Ia larger than 0.5 as shown in Table~\ref{tab:PC_noz}. As shown in Table~\ref{tab:PC_noz}, this stricter cut reduces the number of events while maintaining most of the DES 5-year SNe in \citetalias{Moller:2022}. In Section~\ref{sec:simsalt_contamination}, we estimated the core-collapse contamination of such a photometrically identified sample to be around 6\%. 

This photometric SNe Ia sample is a factor of two larger than the DES 5-year SN Ia sample from \citetalias{Moller:2022} which used redshift information. Our new sample, classified without redshifts, contains $93\%$ of the SNe Ia in \citetalias{Moller:2022}, thus providing reasonably good overlap with less information. Events in \citetalias{Moller:2022} that were not selected when classifying them without redshifts are evenly distributed at all redshifts, with a slight peak around $0.5$, and they have slightly narrower light-curves. While the simultaneous fit is not used for the selection, it provides an indication of the SNIa-likeness of these events. When fitting the light-curves of the lost \citetalias{Moller:2022} SNe, we find systematic offsets in colour, stretch and redshift.

Approximately four percent of the \loose sample have no associated host-galaxy detected with deep photometry in \cite{Wiseman:2020a}. In Section~\ref{sec:fuphost} we discuss further how our selection probes events with fainter hosts than other DES samples which were mostly limited to hosts with $m^{\rm host}_r \leq 24$.

\subsection{High-quality (HQ) sample} \label{sec:hq}

We select a high-quality sample from the $3{,}547$ candidates described in Section~\ref{sec:snn05} by applying cuts on the SNphoto-z and SALT2 parameters fit described in Section~\ref{sec:SALT}. We find that only $2{,}539$ obtain a successful fit. This is due to convergence issues resulting from difficulties to obtain a simultaneous SNphoto-z and SALT2 fit.
 
We select a high-quality (HQ) SN Ia sample shown in Table~\ref{tab:PC_noz} by applying SALT2 cuts introduced in Section~\ref{sec:simsalt_bias}. As the estimation of the SNphoto-z was restricted up to redshift 1.2, we add a cut where SNphoto-z must be below 1.2. We identify $2{,}298$ photometric SNe Ia. This sample is slightly smaller to the expected number of HQ SNe Ia within this redshift range. This small reduction may be due to some issues obtaining SNphoto-z for the SNe Ia consistent with the efficiency estimated in Figure~\ref{fig:eff_noz}. Using simulations, in Section~\ref{sec:simsalt_contamination} we estimate the contamination of a HQ selected sample to be less than 1\%. 

83\% of the DES 5-year SNIa sample in \citetalias{Moller:2022} is also selected in our HQ sample. \citetalias{Moller:2022} SN Ia that were not selected in the HQ sample have differences of up to 0.3 in the SNphoto-zs. We study in more detail the effect of SNphoto-z and the overlap between the samples in Appendix~\ref{appendix:sample_biases_photoz}. Due to this simultaneous fit which offsets significantly the redshift of the event, these SNe Ia have SALT2 parameters that are not compatible with a HQ sample.

We do not find any spectroscopically classified non-Type Ia SN in this HQ sample. We find 7 events in galaxies that have AGN, 5 of them have a separation from the centre of the galaxy $>1''$ and thus are kept. We eliminate two events that are in the centre of the galaxy with an AGN (SNIDs $1303165$, $1257010$).

\begin{figure*}
    \centering
    \includegraphics[width=\textwidth]{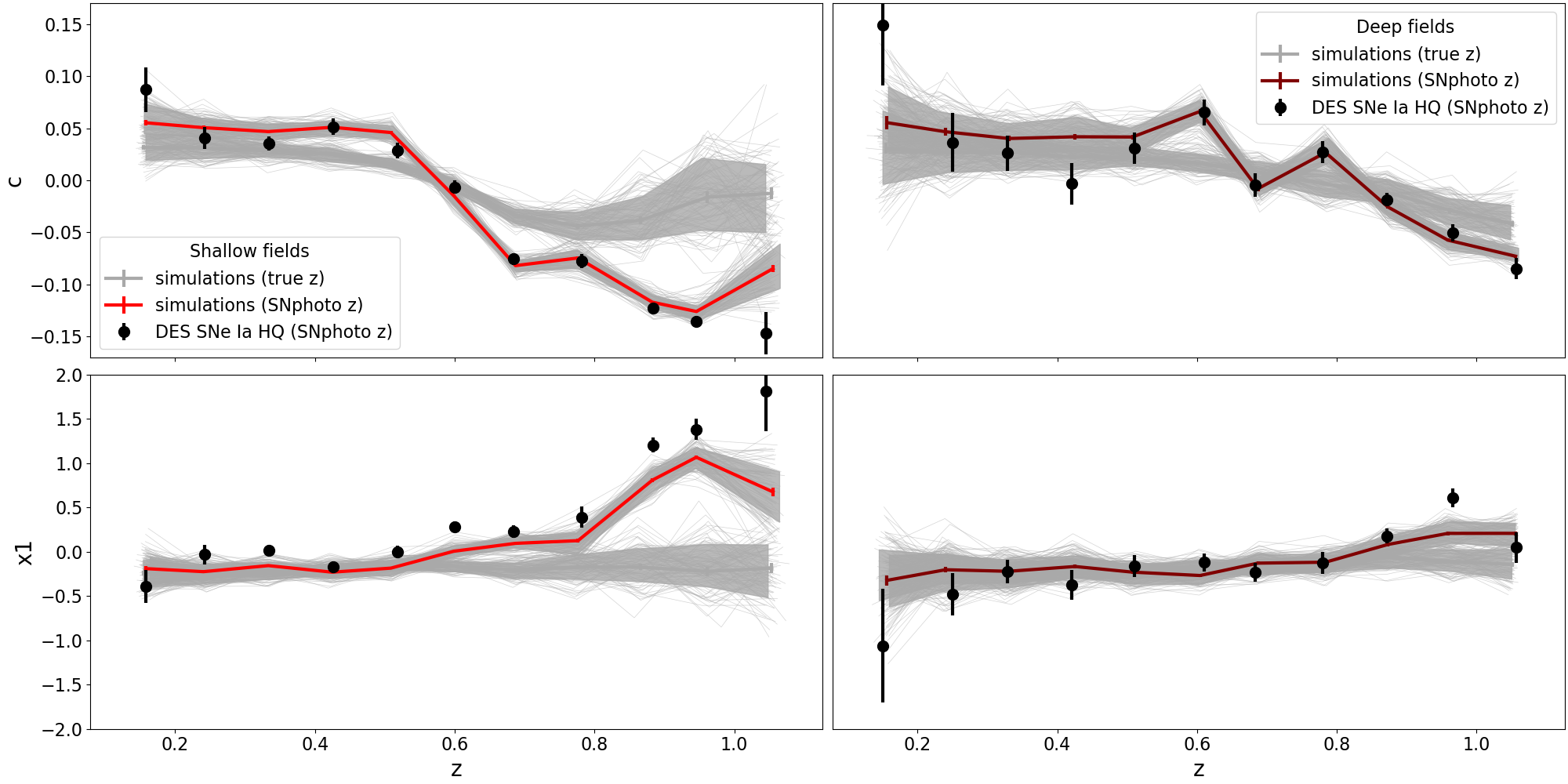}
    \caption{Redshift dependence of SALT2 $c$ and $x1$ for the photometric sample (black markers) and simulated SNe Ia (piecewise curves). We plot SALT2 parameters and SNphoto-z that are simultaneously fit
    for the shallow (red lines in the left-hand panels) and the deep (maroon lines in the right-hand panels) fields. The curves using the simulated (true) redshift are shown in grey. For the simulation, lines are binned averages of the measured parameters. The individual light grey lines represent 150 realisations of the DES-SN 5-year survey and the solid grey filled area covers $68\%$ of these realisations. The mean and the standard deviation of the DES-SN 5-year data are shown using black markers.}
    \label{fig:cx1vs_zHD_noz}
\end{figure*}

\subsection{Sample properties}\label{sec:PC_noz_properties}
In Figure~\ref{fig:hists_noz} we show the redshift and SALT2 measured light-curve parameters for our sample and for simulations as a function of redshift. In the following, true redshift will be the host-galaxy spectroscopic redshift for data and simulated one for simulations; while SNphoto-z will come from the method introduced in Section~\ref{sec:SALT}.

Our photometric sample in the shallow fields agrees better in colour and stretch with simulations using the SNphoto-z, and less with the distribution using parameters derived with the true redshift as shown in the second and third panel in  Figure~\ref{fig:hists_noz}. This reinforces the results from Section~\ref{sec:simsalt_bias} and Figure~\ref{fig:delta_retro} showing that we can simulate and reproduce the biases introduced by the SNphoto-z method. However, for the deep fields we find a better agreement with simulations using the true redshift. This may be due to a reduction of selection effects at high redshift which dominates the shallow fields. We note that for some parameters, such as colour, the distributions with true and SNphoto redshifts are comparable.

We study the redshift evolution of SN Ia light-curve parameters colour and stretch in Figure \ref{fig:cx1vs_zHD_noz}. As the classifier does not use the SNphoto-z nor light-curve parameters, the selected sample is not influenced by the step that estimates these parameters. The differences between simulations in this plot are only due to the values obtained during the SNphoto-z fit.

We find that the data follows the simulation when using SNphoto-z. This suggests that these biases can be reproduced in the simulations. For the deep fields, we observe that the offset from the true z values is coincident with the redshifts were noise starts dominating a band. 

\subsection{Comparing DES SN Ia samples}\label{sec:allDESsamples}

In this Section, we compare differently selected SNe Ia samples from DES: spectroscopically classified, photometrically classified using host-galaxy redshifts \citetalias{Moller:2022}; \cite{DES:5yr,Vincenzi:2024}, and - our current work- a z-free photometrically classified sample. We study SALT2 SN Ia parameters, as well as host-galaxy properties derived in \cite{Wiseman:2020}.

\begin{figure*}
    \includegraphics[width=\textwidth]{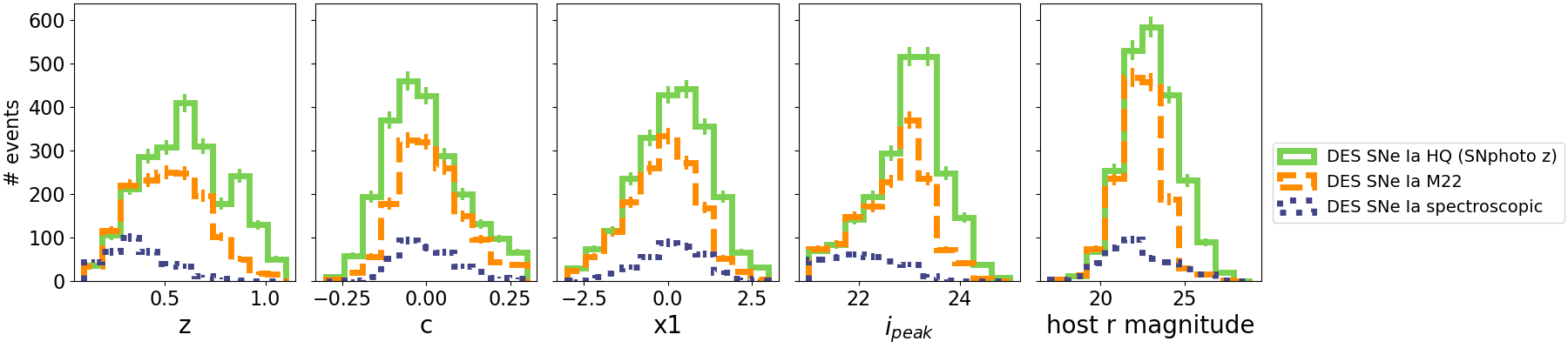}
    \caption{Distributions of redshift, SALT2 $x1$, SALT2 $c$, i-band peak magnitude $i_{\rm peak}$ and host-galaxy r-band magnitude for the HQ sample classified without host information in this work, the photometrically selected SN Ia sample with spectroscopic host-galaxy redshifts \citetalias{Moller:2022} and the spectroscopically classified SNe Ia.}
    \label{fig:allsamples}
\end{figure*}

Host redshifts are only available for a subset of events. We show in Figure~\ref{fig:allsamples} that a sample selected without host or SN redshifts information includes SNe Ia probing a wider range of parameters (e.g. redshift coverage), in greater numbers and in fainter hosts. Our z-free sample also contains SNe Ia that are on average bluer, fainter and with broader light-curves when comparing to spectroscopically classified and photometrically classified with host-galaxy redshift samples.

\begin{figure}
    \includegraphics[width=\columnwidth]{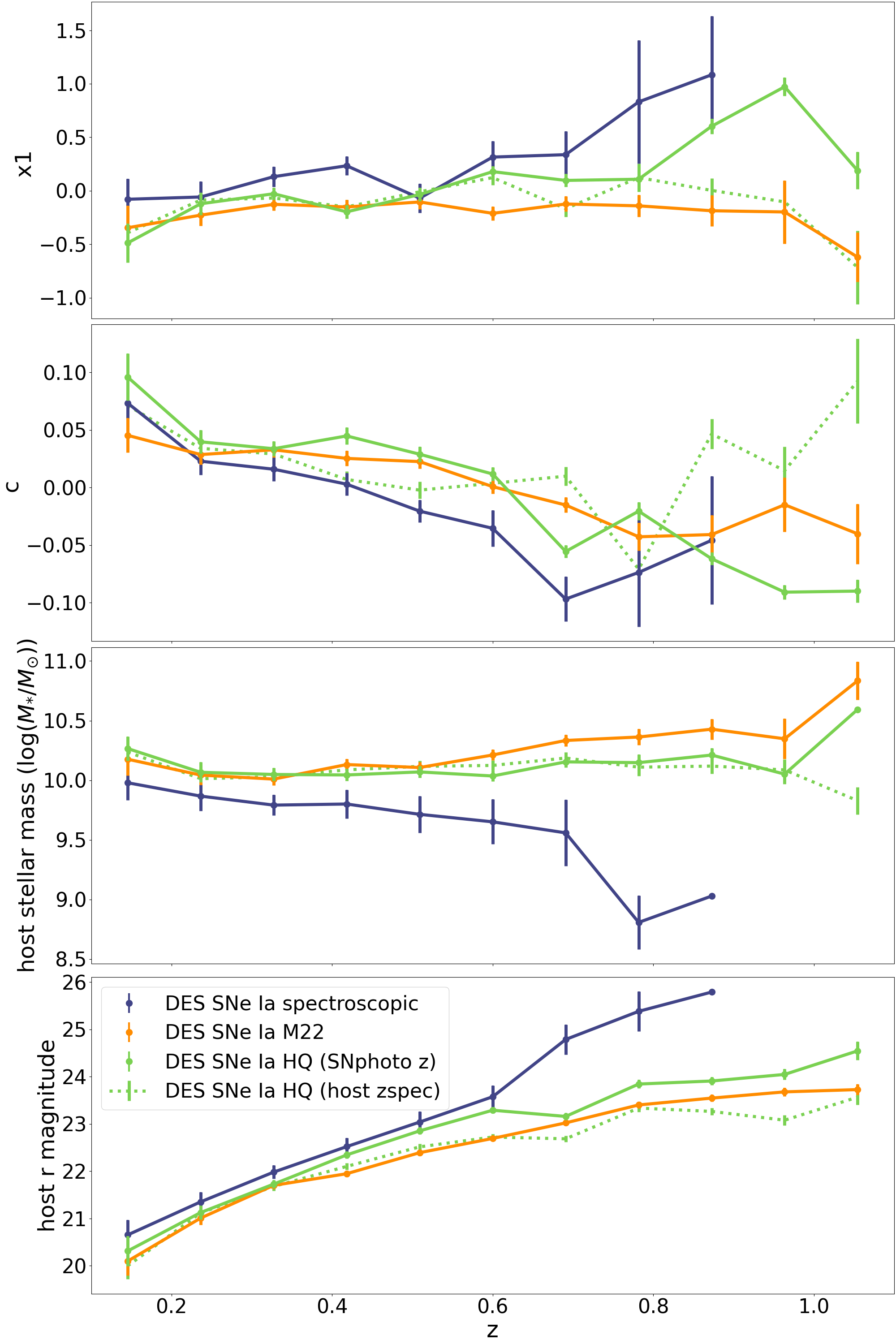}
    \caption{SALT2 stretch and colour, host-galaxy mass and $r$ magnitude as a function of the redshift for the HQ sample classified without host information (green), the photometrically selected SN Ia sample with spectroscopic host-galaxy redshifts (in \citetalias{Moller:2022} in orange) and the spectroscopically classified SNe Ia (in blue). For the sample classified without host information (green) we show two versions: one using SNphoto-z (solid line) computed simultaneously with colour and stretch; and the other using the host-galaxy spectroscopic redshift when available (dotted line). The error bars show the standard error for a given redshift bin. The HQ sample probes SNe Ia in fainter hosts than the \citetalias{Moller:2022} sample at all redshifts as well as lower mass hosts from z>0.4.}
    \label{fig:allsamples_evolz}
\end{figure}

We check the power of our new sample in the context of host-galaxies. For those SNe Ia that have an identified host, we compute their host stellar masses using different sources of redshift. Using SNphoto-z, in Figure~\ref{fig:allsamples_evolz}, we find that the z-free classification includes fainter hosts at all redshifts and with lower masses from z>0.4. We highlight that the z-free sample includes most of M22 plus lower mass galaxies at higher redshifts. We find that the distribution of host-galaxy masses from this sample remains the same at all redshifts below 1 when using masses derived with host-galaxy spectroscopic redshifts or SNphoto-z.

\begin{figure}
    \includegraphics[width=\columnwidth]{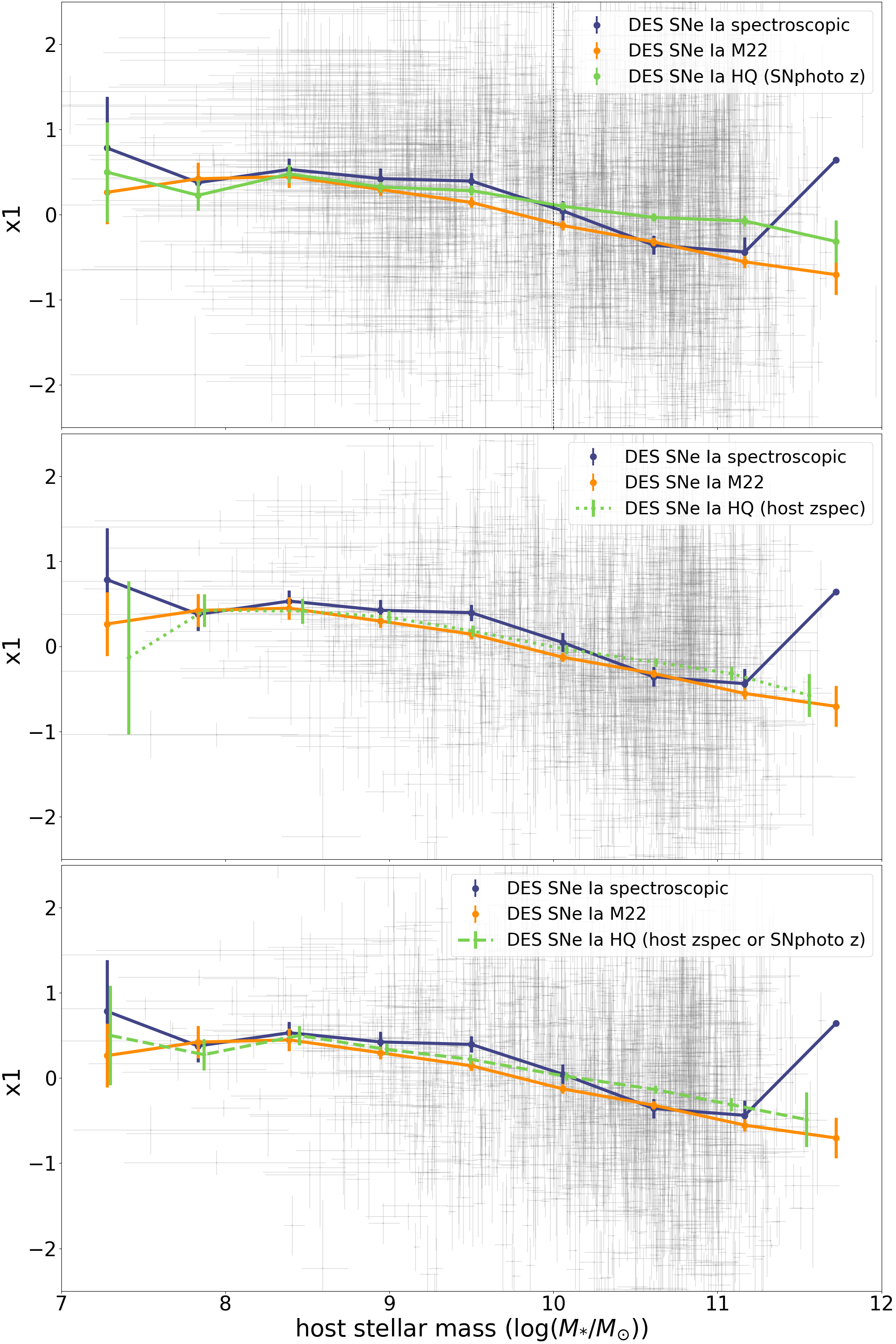}
    \caption{SNIa stretch as a function of host-galaxy mass. In coloured lines we show the median values for the HQ sample classified without host information (green), the photometrically selected SN Ia sample with spectroscopic host-galaxy redshifts (in \citetalias{Moller:2022} in orange) and the spectroscopically classified SNe Ia (in blue). The error bars show the standard error for a given redshift bin. In grey we show each of the measurements for a given SNe Ia in the z-free sample. Each row uses a different redshift for the DES SNe Ia HQ sample and thus its x1 measurement, first row SNphoto-z, second row host-galaxy spectroscopic redshifts if available and third row a mixture of host-galaxy spectroscopic redshift and when not available SNphoto-z. The z-free sample shows for any choice of redshift, a higher stretch at higher mass than the \citetalias{Moller:2022} sample.}
    \label{fig:masstep}
\end{figure}

We further investigate correlations between stretch and host-galaxy mass in Figure~\ref{fig:masstep}. We find that the DES SNe Ia HQ using SNphoto-z have higher stretch at higher masses than other DES samples (1st row). This is also seen even if we restrict to events with host-galaxy spectroscopic redshift in this sample (2nd row) or if we create a "mixed sample" that uses spectroscopic host-galaxy spectroscopic redshifts if available and then SNphoto-z for those without it (3rd row). A detailed study of these correlations is left for future work.

This z-free classified sample will be of value for studying rates, Delay Time Distributions (DTDs), intrinsic populations and in understanding selection biases in our current analyses. However, redshifts are still needed for understanding how these quantities vary through cosmic time. Using the light-curve to estimate redshifts along with light-curve parameters was shown to produce biased estimates. These biases can be reduced by using large redshift bins or by using simulations to correct for the biases. This has been shown in a preliminary analysis with a subset of the DES SN-candidate sample for rates \citep{Lasker:thesis}. For cosmology, another alternative could be to select only candidates in certain types of galaxies such as redMaGiC that can provide accurate host photometric redshifts \citep{Chen:2022} to use for the light-curve fitting or apply SN Ia light-curve redshift driven methods such as the method described in \cite{Qu:2023}.

\begin{figure}
    \centering
    \includegraphics[width=\columnwidth]{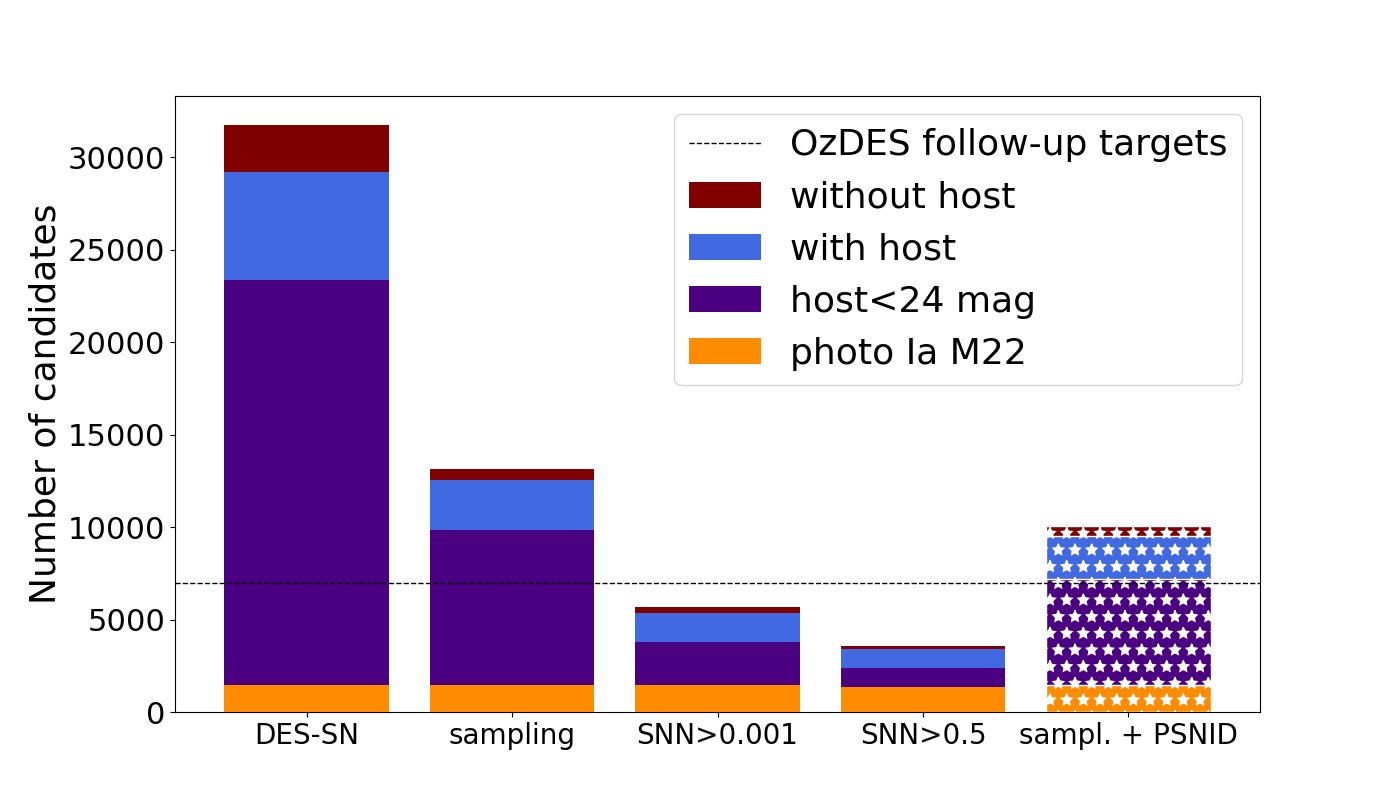}
    \caption{Number of events as a function of cuts applied. The size of the bar includes all events and they are split according to the subsamples (e.g. \citetalias{Moller:2022} photometric SNe Ia in yellow, events with hosts brighter than 24 mag in purple, without host in maroon). From left to right the first four bars represent an additional cut being applied. The right starred bar represents the DES survey follow-up prioritisation strategy: sampling cuts plus a loose cut in PSNID probabilities. We show the number of OzDES follow-up targets as a dashed line.}
    \label{fig:hist_fup_targets}
\end{figure}

\begin{figure}
    \includegraphics[width=1.1\columnwidth]{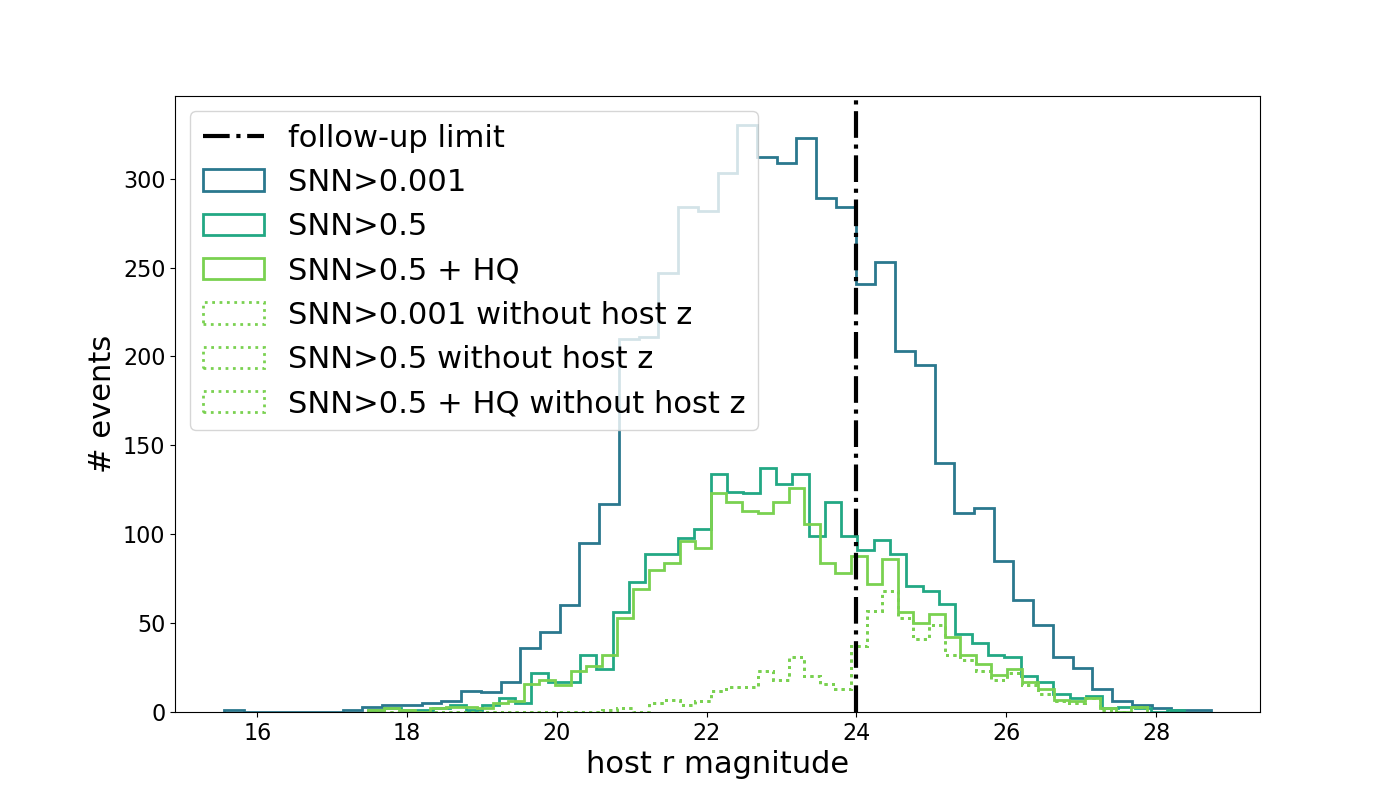}
    \caption{Number of events in the photometrically selected SN Ia sample as a function of host-galaxy $r$ band magnitude. We show samples with different \snn classification scores and a DES cosmology-like cut (solid lines) and those events that had no redshift in the DES database (dotted). The host-galaxy magnitude limit used in OzDES is shown as a vertical line.}
    \label{fig:PC_nohostz}
\end{figure}

\section{Photometric classification for follow-up optimisation}\label{sec:fup}

In this Section we explore how to use photometric classification to optimize spectroscopic follow-up of host-galaxies (Section~\ref{sec:fuphost}) and SNe while still bright enough to observe and preferably before maximum light (Section~\ref{sec:early}).

\subsection{Follow-up of host-galaxies} \label{sec:fuphost}

Host-galaxy follow-up provides accurate redshifts, which are needed for the Hubble diagram and thus cosmology. As spectroscopic resources are scarce, prioritization of potential SN Ia host-galaxies is necessary for spectroscopic follow-up programmes. 

The Australian Dark Energy Survey (OzDES) provided multi-object fibre spectroscopy for the Dark Energy Survey using the 2dF fibre positioner and AAOmega spectrograph on the 3.9-m Anglo-Australian Telescope \citep{Yuan:2015,Childress:2017,Lidman:2020}. OzDES targeted a wide range of sources over the six years, with active transients, AGN, and host-galaxies with $r< 24$ having the highest priority and occupying most of the fibres.

For DES, OzDES targeted $8{,}666$ candidate SN hosts and obtained redshifts for $6{,}391$ of these galaxies \citep{Lidman:2020}. OzDES targets were selected from $31{,}636$ DES SN candidates by prioritising those with a high probability of being SNe Ia from fits with the Photometric Supernova IDentification software \citep[PSNID]{Sako:2011} and selecting hosts mostly with ${r}<24$.

In this Section, we explore using \snn probabilities for host-galaxy spectroscopic follow-up prioritisation. This will be crucial for future surveys such as Rubin LSST and its follow-up programme the Time-Domain Extragalactic Survey \citep[TiDES;][]{Swann:2019} on the 4-metre Multi-Object Spectrograph Telescope (4MOST).

\subsection*{SNN>0.001}
We explore the use of a loose cut in SN Ia probability for identifying potential host-galaxies. We apply an SNN>0.001 cut after pre-processing cuts in Section~\ref{sec:preprocessing}. From the pre-processed $13{,}118$ candidates, this cut reduces significantly the sample to $5{,}676$ as shown in the fourth row of Table~\ref{tab:PC_noz} and in the third bar of Figure~\ref{fig:hist_fup_targets}
while keeping all events in the \citetalias{Moller:2022} sample except 3 (shown in orange in Figure~\ref{fig:hist_fup_targets}). 

To estimate the performance of this selection method we apply the same SNN>0.001 cut to DES SN simulations. We recover almost all of the simulated SNe Ia, obtaining an efficiency of $99.996 \pm 0.001 \%$. As this is a loose cut, we find that the purity of the sample is only $88.68 \pm 0.06 \%$. For one realisation of the DES survey, with $4{,}962$ type Ia and $3{,}466$ core-collapse SNe, this cut selects all the type Ia and $613$ core-collapse SNe. We highlight, that this simulation does not include other types of transients and spurious detections which may contaminate the candidates.

We now compare this loose \ssnn cut with respect to the method used during the DES survey to prioritise potential SNe host-galaxies. During the DES survey, a loose PSNID probability cut was used to select candidates. In Figure~\ref{fig:hist_fup_targets}, we find that a loose SNN probability cut (third bar) reduces the follow-up sample from PSNID by a factor of two while maintaining the number of DES 5-year SNe Ia \citetalias{Moller:2022} (yellow bar).

Future surveys, such as Rubin LSST, will not only require accurate selection of candidates, but also scalable methods to address the big data volumes. \snn has been shown to be scalable, classifying thousands of light-curves per second.

\subsection*{SNN>0.5}
We now explore whether a tighter probability cut provides a good sample for host-galaxy follow-up. 

Using simulations, we estimate that such a cut would recover $4{,}883$ from $4{,}962$ type Ia and select $128$ from $3{,}466$ non Ia SNe. This represents an efficiency of $98.36 \pm 0.01 \%$ and purity of  $97.297 \pm 0.004 \%$. As in the previous Section, these simulations only indicate the performance on SN light-curves while the data may include other transients and spurious detections.

We apply a \loose threshold as in Section~\ref{sec:snn05} to the DES SN-candidates, finding that there is a significant reduction on the number of follow-up candidates, while maintaining the number SNe Ia. From these $3{,}547$ host-galaxy follow-up candidates, $1{,}441$ have no spectroscopic redshift from DES follow-up programmes.

As shown in Figure~\ref{fig:PC_nohostz}, most of the host-galaxies without redshift are faint. In the context of DES, if we select those events in host-galaxies with a limiting magnitude similar to that for spectroscopic follow-up at the OzDES programme, $m^{\rm host}_r \leq 24$, we obtain $2{,}394$ potential follow-up host-galaxies. Most of these galaxies were followed-up and a redshift was acquired. The majority of hosts without redshifts come from SN candidates in the last two years of the survey, which had less time to be followed-up and thus resulted in shorter integration times. These selection effects were modelled by \cite{Vincenzi:2022}.

This method provides potential prioritisation for follow-up galaxies to extend the DES 5-year sample with 447 new events with hosts within the magnitude limits of the AAT and the OzDES programme.

\subsection{Early classification for live SN follow-up}\label{sec:early}
In this Section, we explore the early identification of candidates for SN spectroscopic follow-up optimisation. This identification is done with partial light-curves, preferably before maximum brightness.

DES light-curves are preprocessed using the following cuts:
\begin{itemize}
    \item Artifacts are rejected using the transient\_status flag as in \cite{Smith:2018}.
    \item We eliminate photometry that has been flagged as flawed using bitmap flags from {\sc Source Extractor} \citep{Bertin:1996}.
\end{itemize}

To trigger follow-up, a sequence of detections must be identified. The DES trigger required at least one detection on 2 separate nights. To verify its performance, we select photometric measurements (i) within a time-window of 7 days before to 20 days after the DES-like trigger and (ii) within a time-window of 30 days before the observed peak and the observed peak. We apply $SNN>0.5$ classification threshold to select candidates for follow-up as shown in Table~\ref{tab:early_class}.

We find that the median number of detections per SN in all bands for early classification using the DES-like trigger and peak selection methods respectively are: (i) $7\pm4$ and (ii) $6\pm5$ (errors are given by one standard deviation for all SNe). 

We compare our selection for potential live SN follow-up with the OzDES strategy for a magnitude limited sample. OzDES obtained 1460 spectra of live-transients prioritising events that were brighter than $22.7$. As shown in Table~\ref{tab:early_class}, for candidates with any band magnitude $<22.7$ we find that \ssnn reduces the number of potential follow-up candidates by more than a factor of 3, maintaining most of the SNe Ia.

Interestingly, \snn is able to eliminate a large fraction of multiseason (e.g. AGN) events. These events were not part of the training set and this indicates that the classification is robust to out-of-distribution events.

\begin{table}
    \caption{Selection of targets for spectroscopic follow-up. The first two rows show the number of events selected from their partial light-curves using photometry -7<DES-like trigger<20 days. The following two rows show the same statistics but for light-curves selected within a time-window of -30<peak<1 and then for -7<LSST-like trigger<20. For all cases, we only include events with peak magnitudes brighter than 22.7 in any band, which was the OzDES limiting magnitude for live transient follow-up. }.
    \label{tab:early_class}
\begin{tabular}{lrrrrr}
cut &  total &  specIa  &  M22 &  spec nonIa &  multiseason \\
\hline
\multicolumn {6} {c}{ \textit{-7<DES-like trigger<20}}\\
\hline
-7<DES<20 &               3250 &                 336 &              776 &                120 &                      230 \\
SNN>0.5 &               1288 &                 294 &              687 &                  4 &                       18 \\

\hline 
\multicolumn {6} {c}{ \textit{-30<peak<1}}\\
\hline
-30<peak<1 &               5702 &                 359 &              810 &                144 &                      622 \\
SNN>0.5 &               1428 &                 305 &              683 &                  4 &                       19 \\
\hline
\multicolumn {6} {c}{ \textit{-7<LSST-like trigger<20}}\\
\hline
-7<LSST<20 &               3327 &                 296 &              689 &                 95 &                      219 \\
 SNN>0.5 &               1305 &                 264 &              618 &                  4 &                       28 \\

\end{tabular}
\end{table}

\section{Prospectives for Rubin and 4MOST}\label{sec:rubin}

The Vera C. Rubin Observatory is expected to obtain up to 10 million detections (alerts) of transients every night during the 10-year Legacy Survey of Space and Time  \citep[LSST;][]{ldm612}. There is the potential of discovering hundreds of thousands supernovae for cosmology and astrophysical studies \citep{LSST:2009,TVS:2022}. This is several orders of magnitudes larger than DES. Rubin LSST will provide multi-band light-curves for all these transients. The 4MOST Time-Domain Extragalactic Survey \citep[TiDES;][]{Swann:2019} will provide a large fraction of follow-up for host-galaxies and live transients with spectroscopy.

Given the sheer volume of data from LSST, it will be crucial to optimise resources for the spectroscopic follow-up of hosts-galaxies and live supernovae.

TiDES is expected to obtain host-galaxy redshifts for 50,000 SNe Ia up to redshift of 1. In Section~\ref{sec:fuphost} we have shown that \snn can drastically reduce the number of candidates sent for host-galaxy follow up while maintaining most of the SNe Ia in the sample. 

For follow-up of live transients, LSST will emit an alert when a detection occurs with S/N>5. These alerts will be received by Rubin Community brokers \citep[e.g. {\sc Fink},][]{Moller:2021}. In Table~\ref{tab:early_class} we show the effect of using a single detection for the DES data to select early SNe. In the following, we explore the adequacy of a single LSST-like trigger and then explore a follow-up similar in magnitude depth as TiDES.

\subsection*{Is a LSST-like trigger a good indicator for a real event?} 

We now test an LSST-like trigger, where only one detection with S/N>5 is required. Intuitively, the LSST-like trigger time should be within a month of the observed peak for SNe. We check whether the LSST-like trigger is within 30 days of the observed peak finding only 85\% for the DES 5-year photometric SN Ia sample (in \citetalias{Moller:2022}) and 81\% for the spectroscopically classified SNe Ia satisfy this condition.

\begin{figure*}
\centering
    \includegraphics[width=\textwidth]{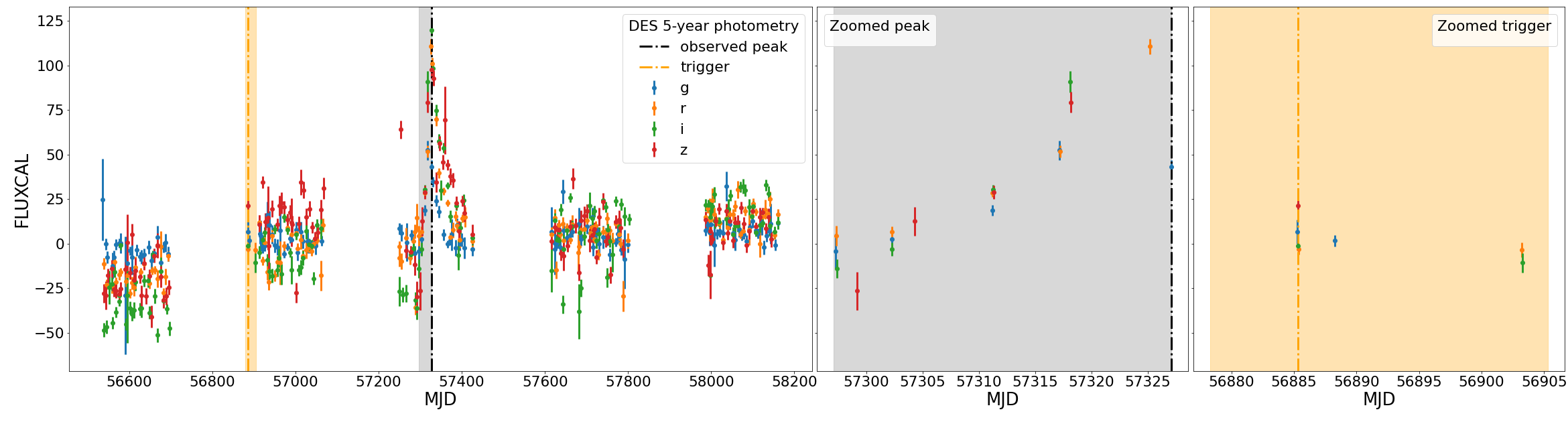}
    \caption{The full five-year light-curve of DES15C3lvt in the left panel. The measured flux in $g, r, i, z$ bands, plotted against Modified Julien Date (MJD). In the centre and right panels we show the photometry used for classification before observed peak (grey) and around trigger (orange). For this event the trigger, or first detection, is far away from the SN and was probably due to noise.}
    \label{fig:lc_peak_trigger}
\end{figure*}

These results show that a LSST-like trigger is not necessarily a robust indicator of the start of the SN event for large surveys. An example of a SN with a trigger in a different year than the event is shown in Figure~\ref{fig:lc_peak_trigger}.

Importantly, using a single detection as a criterion for follow-up will not optimise our scarce follow-up resources. A larger fraction of LSST-like triggers when compared to a DES-like trigger will not correspond to a SN-like event. To reduce the number of spurious detections it will be necessary to increase the number of detections necessary for follow-up and monitor whether the light-curve is rising in brightness together with a classifier (e.g. \cite{Leoni:2022}) or to implement a requirement for a second detection within 30 days as in DES.

\subsection*{TiDES-like selection}\label{sec:tides}
Simulations predict that TiDES will be able to classify live transients as faint as $r_{\rm mag}\approx 22.5$ \citep{Swann:2019}. In this Section, we discuss the early classification of transients in the DES survey as a precursor for the Rubin LSST SN sample.

The main contamination is multiseason events identified a posteriori by their detection over multiple seasons. For Rubin, it could be beneficial to incorporate AGN models into the training set to reduce this contaminant or to filter out these events using pre-existing photometry if this photometry is available. For example, much of the area that LSST will cover has imaging data with DECam.

Importantly, for DES we found that the LSST-like trigger can sometimes occur much earlier than the SN event as a result of noise fluctuation or subtraction artifact. This can be an issue if classification is restricted to a small time window around trigger. Thus, to avoid losing potential SN, a DES-like trigger could be applied or an strategy could be applied where detections are classified regardless of the trigger time with algorithms that can classify SNe at any time step. To increase purity, additional requirements can be included such as a second detection night or rising light-curves.

In this work we use data from DES as a precursor for Rubin LSST. Using LSST simulations, other works have explored: the optimisation of the 4MOST follow-up strategy \cite{Carrick:2021} and the rate of recovery of SNe Ia using \snn (Petreca et al. in prep).

\section{Conclusions}

In this work we photometrically classified SNe Ia from the 5-year DES survey data using only light-curves and the framework \snnns. Our goal was to classify detected SNe Ia regardless of whether their hosts could be identified. In anticipation of future surveys, we also explore the use of \snn to optimise follow-up resources for host-galaxies and live SNe.

From the DES 5-year data we obtain $3{,}547$ SNe Ia, photometrically classified without using any redshift information. This sample doubles the DES 5-year sample classified with host-galaxy redshifts in \cite{Moller:2022,DES:5yr, Vincenzi:2024} and contains SNe Ia in faint galaxies. 

To obtain a high-quality SN Ia sample, we first estimate redshifts from the SN light-curves using the SNphoto-z method \citep{Kessler:2010FittedPhotoZ}. We then use the redshifts and light-curve parameters to restrict our sample to $2,298$ high-quality SNe Ia. This is consistent with the estimated number of well measured SNe Ia in DES according to simulations.

We find that this HQ sample contains $83\%$ of the previous SNe Ia sample classified with host-redshifts in \citetalias{Moller:2022}. Most of the \citetalias{Moller:2022} SNe Ia are lost due to lack of convergence of the SNphoto-z. If new host-galaxy photo-z are available, combining the SNphoto-z method with a host-galaxy photo-z prior is expected to significantly improve photo-z estimates and the fitting efficiency \citep{Mitra:2023}.

We also find that there are structured offsets between the estimates of SNphoto-z and SALT2 parameters with respect to the true values in simulations. However, we find potential for using this sample with SNphoto-zs for analyses in the deep fields or in analyses that require a binning over redshift or other parameters.

Future surveys such as Rubin LSST will continue to detect more SNe than it is possible to follow-up spectroscopically both for classification and host-galaxy redshift acquisition. In this work, we also show that \snn is more effective than previous methods at reducing the number of candidates for host-galaxy (four times) and live SN  (three times) follow-up while maintaining the number of SNe Ia. Importantly, it significantly reduces contaminants such as AGN which were not used for training as they are challenging to simulate.  

We use our DES results to examine potential challenges and solutions for Rubin LSST and the spectroscopic time domain follow-up programme 4MOST TiDES. In particular for live SN follow-up we find that using an LSST-like trigger (only 1 detection SNR>5) yields a large number of triggers not coincident with real SNe detections. We find that an alternative to improve triggering is to use a DES-like trigger to define the time region for classification. 

In this work we have identified most of the expected SNe Ia in the DES dataset. When compared to other DES SN Ia samples both the spectroscopically classified and the photometrically classified using host-galaxy redshifts in \citetalias{Moller:2022}, we find that we are probing higher redshift, fainter, bluer and higher stretch SNe Ia populations. For those SNe Ia in this sample with an identified host, we find that we are probing lower host-galaxy masses at high-redshifts and at higher host masses we are obtaining higher stretch SNe Ia.

A purely light-curve classified SN Ia sample, such as the one in this work, harnesses the power of large surveys such as DES. These large statistical sample, has the potential to further shed light on questions about SNe Ia diversity and environments.

\section*{Acknowledgements}

AM is supported by the ARC Discovery Early Career Researcher Award (DECRA) project number DE230100055. LG acknowledges financial support from the Spanish Ministerio de Ciencia e Innovaci\'on (MCIN) and the Agencia Estatal de Investigaci\'on (AEI) 10.13039/501100011033 under the PID2020-115253GA-I00 HOSTFLOWS project, from Centro Superior de Investigaciones Cient\'ificas (CSIC) under the PIE project 20215AT016 and the program Unidad de Excelencia Mar\'ia de Maeztu CEX2020-001058-M, and from the Departament de Recerca i Universitats de la Generalitat de Catalunya through the 2021-SGR-01270 grant.

Author contributions. AM performed the analysis and wrote the manuscript. The top-tier authors aided in the interpretation of the analysis and: PW constructed the host-galaxy catalogue; MSmith, CL and TD were internal reviewers, collected and reduced data; MSullivan computed host-galaxy masses; RK provided advice on SNphoto-z and simulations; MSako was internal reviewer. The following authors contributed to the analysis of the DES5YR dataset: LG, JL, RN, BS, BT.
The remaining authors have made contributions to this paper that include, but are not limited to, the construction of DECam and other aspects of collecting the data; data processing and calibration; developing broadly used methods, codes, and simulations; running the pipelines and validation tests; and promoting the science analysis.

This paper has gone through internal review by the DES collaboration. Funding for the DES Projects has been provided by the U.S. Department of Energy, the U.S. National Science Foundation, the Ministry of Science and Education of Spain, the Science and Technology Facilities Council of the United Kingdom, the Higher Education Funding Council for England, the National Center for Supercomputing Applications at the University of Illinois at Urbana-Champaign, the Kavli Institute of Cosmological Physics at the University of Chicago, the Center for Cosmology and Astro-Particle Physics at the Ohio State University, the Mitchell Institute for Fundamental Physics and Astronomy at Texas A\&M University, Financiadora de Estudos e Projetos, Fundação Carlos Chagas Filho de Amparo à Pesquisa do Estado do Rio de Janeiro, Conselho Nacional de Desenvolvimento Científico e Tecnológico and the Ministério da Ciência, Tecnologia e Inovação, the Deutsche Forschungsgemeinschaft and the Collaborating Institutions in the Dark Energy Survey.
The Collaborating Institutions are Argonne National Laboratory, the University of California at Santa Cruz, the University of Cambridge, Centro de Investigaciones Energéticas, Medioambientales y Tecnológicas-Madrid, the University of Chicago, University College London, the DES-Brazil Consortium, the University of Edinburgh, the Eidgenössische Technische Hochschule (ETH) Zürich, Fermi National Accelerator Laboratory, the University of Illinois at Urbana-Champaign, the Institut de Ciències de l’Espai (IEEC/CSIC), the Institut de Física d’Altes Energies, Lawrence Berkeley National Laboratory, the Ludwig-Maximilians Universität München and the associated Excellence Cluster Universe, the University of Michigan, NFS’s NOIRLab, the University of Nottingham, The Ohio State University, the University of Pennsylvania, the University of Portsmouth, SLAC National Accelerator Laboratory, Stanford University, the University of Sussex, Texas A\&M University, and the OzDES Membership Consortium.

Based in part on observations at Cerro Tololo Inter-American Observatory at NSF’s NOIRLab (NOIRLab Prop. ID 2012B-0001; PI: J. Frieman), which is managed by the Association of Universities for Research in Astronomy (AURA) under a cooperative agreement with the National Science Foundation.

The DES data management system is supported by the National Science Foundation under Grant Numbers AST-1138766 and AST-1536171. The DES participants from Spanish institutions are partially supported by MICINN under grants ESP2017-89838, PGC2018-094773, PGC2018-102021, SEV-2016-0588, SEV-2016-0597, and MDM-2015-0509, some of which include ERDF funds from the European Union. IFAE is partially funded by the CERCA program of the Generalitat de Catalunya. Research leading to these results has received funding from the European Research Council under the European Union’s Seventh Framework Program (FP7/2007-2013) including ERC grant agreements 240672, 291329, and 306478. We acknowledge support from the Brazilian Instituto Nacional de Ciência e Tecnologia (INCT) do e-Universo (CNPq grant 465376/2014- 2).

This work was completed in part with Midway resources provided by the University of Chicago’s Research Computing Center. 

This work makes use of data acquired at the Anglo-Australian Telescope, under program A/2013B/012. We acknowledge the traditional owners of the land on which the AAT stands, the Gamilaraay people, and pay our respects to elders past and present.

\section*{Data Availability}

For reproducibility we provide: (i) the SNANA and/or Pippin configurations to reproduce the simulations in this paper and \citep{Moller:2022} at \url{https://github.com/anaismoller/DES5YR_SNeIa_hostz}, (ii) analysis code in python to reproduce plots and results at \url{https://github.com/anaismoller/DES5YR_SNeIa_nohost}. The DES SN-candidate data will be released by the Collaboration at a later stage.



\bibliographystyle{mnras}
\bibliography{ref} 

\begin{thebibliography}{}
\makeatletter
\relax
\def\mn@urlcharsother{\let\do\@makeother \do\$\do\&\do\#\do\^\do\_\do\%\do\~}
\def\mn@doi{\begingroup\mn@urlcharsother \@ifnextchar [ {\mn@doi@} {\mn@doi@[]}}
\def\mn@doi@[#1]#2{\def\@tempa{#1}\ifx\@tempa\@empty \href {http://dx.doi.org/#2} {doi:#2}\else \href {http://dx.doi.org/#2} {#1}\fi \endgroup}
\def\mn@eprint#1#2{\mn@eprint@#1:#2::\@nil}
\def\mn@eprint@arXiv#1{\href {http://arxiv.org/abs/#1} {{\tt arXiv:#1}}}
\def\mn@eprint@dblp#1{\href {http://dblp.uni-trier.de/rec/bibtex/#1.xml} {dblp:#1}}
\def\mn@eprint@#1:#2:#3:#4\@nil{\def\@tempa {#1}\def\@tempb {#2}\def\@tempc {#3}\ifx \@tempc \@empty \let \@tempc \@tempb \let \@tempb \@tempa \fi \ifx \@tempb \@empty \def\@tempb {arXiv}\fi \@ifundefined {mn@eprint@\@tempb}{\@tempb:\@tempc}{\expandafter \expandafter \csname mn@eprint@\@tempb\endcsname \expandafter{\@tempc}}}

\bibitem[\protect\citeauthoryear{{Abbott} et~al.,}{{Abbott} et~al.}{2018}]{Abbott:2018}
{Abbott} T.~M.~C.,  et~al., 2018, \mn@doi [\apjs] {10.3847/1538-4365/aae9f0}, \href {https://ui.adsabs.harvard.edu/abs/2018ApJS..239...18A} {239, 18}

\bibitem[\protect\citeauthoryear{{Abbott} et~al.,}{{Abbott} et~al.}{2019}]{DES:3yr}
{Abbott} T.~M.~C.,  et~al., 2019, \mn@doi [\apjl] {10.3847/2041-8213/ab04fa}, \href {https://ui.adsabs.harvard.edu/abs/2019ApJ...872L..30A} {872, L30}

\bibitem[\protect\citeauthoryear{{Bazin} et~al.,}{{Bazin} et~al.}{2011}]{Bazin:2011}
{Bazin} G.,  et~al., 2011, \mn@doi [\aap] {10.1051/0004-6361/201116898}, \href {https://ui.adsabs.harvard.edu/abs/2011A&A...534A..43B} {534, A43}

\bibitem[\protect\citeauthoryear{Bellm et~al.,}{Bellm et~al.}{2018}]{Bellm_2018}
Bellm E.~C.,  et~al., 2018, \mn@doi [Publications of the Astronomical Society of the Pacific] {10.1088/1538-3873/aaecbe}, 131, 018002

\bibitem[\protect\citeauthoryear{Bellm et~al.,}{Bellm et~al.}{2019}]{ldm612}
Bellm E.,  et~al., 2019, {LDM-612, Plans and Policies for LSST Alert Distribution}, \url{https://ls.st/ldm-612}

\bibitem[\protect\citeauthoryear{Bernstein et~al.}{Bernstein et~al.}{2012}]{Bernstein:2011}
Bernstein J.~P.,  et~al., 2012, \mn@doi [Astrophys. J.] {10.1088/0004-637X/753/2/152}, 753, 152

\bibitem[\protect\citeauthoryear{{Bertin} \& {Arnouts}}{{Bertin} \& {Arnouts}}{1996}]{Bertin:1996}
{Bertin} E.,  {Arnouts} S.,  1996, \mn@doi [\aaps] {10.1051/aas:1996164}, \href {https://ui.adsabs.harvard.edu/abs/1996A&AS..117..393B} {117, 393}

\bibitem[\protect\citeauthoryear{{Betoule} et~al.,}{{Betoule} et~al.}{2014}]{Betoule:2014}
{Betoule} M.,  et~al., 2014, \mn@doi [\aap] {10.1051/0004-6361/201423413}, \href {https://ui.adsabs.harvard.edu/abs/2014A&A...568A..22B} {568, A22}

\bibitem[\protect\citeauthoryear{{Boone}}{{Boone}}{2021}]{Boone:2021}
{Boone} K.,  2021, \mn@doi [\aj] {10.3847/1538-3881/ac2a2d}, \href {https://ui.adsabs.harvard.edu/abs/2021AJ....162..275B} {162, 275}

\bibitem[\protect\citeauthoryear{{Brout} et~al.,}{{Brout} et~al.}{2022}]{Brout:2022}
{Brout} D.,  et~al., 2022, \mn@doi [\apj] {10.3847/1538-4357/ac8e04}, \href {https://ui.adsabs.harvard.edu/abs/2022ApJ...938..110B} {938, 110}

\bibitem[\protect\citeauthoryear{{Carrick}, {Hook}, {Swann}, {Boone}, {Frohmaier}, {Kim}, {Sullivan}  \& {LSST Dark Energy Science Collaboration}}{{Carrick} et~al.}{2021}]{Carrick:2021}
{Carrick} J.~E.,  {Hook} I.~M.,  {Swann} E.,  {Boone} K.,  {Frohmaier} C.,  {Kim} A.~G.,  {Sullivan} M.,   {LSST Dark Energy Science Collaboration} 2021, \mn@doi [\mnras] {10.1093/mnras/stab2343}, \href {https://ui.adsabs.harvard.edu/abs/2021MNRAS.508....1C} {508, 1}

\bibitem[\protect\citeauthoryear{{Chen} et~al.,}{{Chen} et~al.}{2022}]{Chen:2022}
{Chen} R.,  et~al., 2022, \mn@doi [\apj] {10.3847/1538-4357/ac8b82}, \href {https://ui.adsabs.harvard.edu/abs/2022ApJ...938...62C} {938, 62}

\bibitem[\protect\citeauthoryear{{Chen} et~al.,}{{Chen} et~al.}{2024}]{Chen:2024}
{Chen} R.,  et~al., 2024, \mn@doi [arXiv e-prints] {10.48550/arXiv.2407.16744}, \href {https://ui.adsabs.harvard.edu/abs/2024arXiv240716744C} {p. arXiv:2407.16744}

\bibitem[\protect\citeauthoryear{{Childress} et~al.,}{{Childress} et~al.}{2017}]{Childress:2017}
{Childress} M.~J.,  et~al., 2017, \mn@doi [\mnras] {10.1093/mnras/stx1872}, \href {https://ui.adsabs.harvard.edu/abs/2017MNRAS.472..273C} {472, 273}

\bibitem[\protect\citeauthoryear{{Contreras} et~al.,}{{Contreras} et~al.}{2010}]{Contreras:2010}
{Contreras} C.,  et~al., 2010, \mn@doi [\aj] {10.1088/0004-6256/139/2/519}, \href {https://ui.adsabs.harvard.edu/abs/2010AJ....139..519C} {139, 519}

\bibitem[\protect\citeauthoryear{{DES Collaboration}}{{DES Collaboration}}{2024}]{DES:5yr}
{DES Collaboration} 2024, \mn@doi [arXiv e-prints] {10.48550/arXiv.2401.02929}, \href {https://ui.adsabs.harvard.edu/abs/2024arXiv240102929D} {p. arXiv:2401.02929}

\bibitem[\protect\citeauthoryear{{Doi} et~al.,}{{Doi} et~al.}{2010}]{Doi:2010}
{Doi} M.,  et~al., 2010, \mn@doi [\aj] {10.1088/0004-6256/139/4/1628}, \href {https://ui.adsabs.harvard.edu/abs/2010AJ....139.1628D} {139, 1628}

\bibitem[\protect\citeauthoryear{{Flaugher} et~al.,}{{Flaugher} et~al.}{2015}]{Flaugher:2015}
{Flaugher} B.,  et~al., 2015, \mn@doi [\aj] {10.1088/0004-6256/150/5/150}, \href {https://ui.adsabs.harvard.edu/abs/2015AJ....150..150F} {150, 150}

\bibitem[\protect\citeauthoryear{{Fraga} et~al.,}{{Fraga} et~al.}{2024}]{Fraga:2024}
{Fraga} B.~M.~O.,  et~al., 2024, \mn@doi [arXiv e-prints] {10.48550/arXiv.2404.08798}, \href {https://ui.adsabs.harvard.edu/abs/2024arXiv240408798F} {p. arXiv:2404.08798}

\bibitem[\protect\citeauthoryear{{Frohmaier} et~al.,}{{Frohmaier} et~al.}{2019}]{Frohmaier:2019}
{Frohmaier} C.,  et~al., 2019, \mn@doi [\mnras] {10.1093/mnras/stz807}, \href {https://ui.adsabs.harvard.edu/abs/2019MNRAS.486.2308F} {486, 2308}

\bibitem[\protect\citeauthoryear{Gagliano, Contardo, Foreman-Mackey, Malz  \& Aleo}{Gagliano et~al.}{2023}]{Gagliano:2023}
Gagliano A.,  Contardo G.,  Foreman-Mackey D.,  Malz A.~I.,   Aleo P.~D.,  2023, \mn@doi [The Astrophysical Journal] {10.3847/1538-4357/ace326}, 954, 6

\bibitem[\protect\citeauthoryear{{Goldstein} et~al.,}{{Goldstein} et~al.}{2015}]{Goldstein:2015}
{Goldstein} D.~A.,  et~al., 2015, \mn@doi [\aj] {10.1088/0004-6256/150/3/82}, \href {https://ui.adsabs.harvard.edu/abs/2015AJ....150...82G} {150, 82}

\bibitem[\protect\citeauthoryear{{Guy} et~al.,}{{Guy} et~al.}{2007}]{Guy:2007}
{Guy} J.,  et~al., 2007, \mn@doi [\aap] {10.1051/0004-6361:20066930}, \href {https://ui.adsabs.harvard.edu/abs/2007A&A...466...11G} {466, 11}

\bibitem[\protect\citeauthoryear{{Hambleton} et~al.,}{{Hambleton} et~al.}{2023}]{TVS:2022}
{Hambleton} K.~M.,  et~al., 2023, \mn@doi [\pasp] {10.1088/1538-3873/acdb9a}, \href {https://ui.adsabs.harvard.edu/abs/2023PASP..135j5002H} {135, 105002}

\bibitem[\protect\citeauthoryear{{Hicken} et~al.,}{{Hicken} et~al.}{2009}]{Hicken:2009}
{Hicken} M.,  et~al., 2009, \mn@doi [\apj] {10.1088/0004-637X/700/1/331}, \href {https://ui.adsabs.harvard.edu/abs/2009ApJ...700..331H} {700, 331}

\bibitem[\protect\citeauthoryear{{Hinton} \& {Brout}}{{Hinton} \& {Brout}}{2020}]{Hinton:2020}
{Hinton} S.,  {Brout} D.,  2020, \mn@doi [The Journal of Open Source Software] {10.21105/joss.02122}, \href {https://ui.adsabs.harvard.edu/abs/2020JOSS....5.2122H} {5, 2122}

\bibitem[\protect\citeauthoryear{{Hlozek} et~al.,}{{Hlozek} et~al.}{2012}]{Hlozek:2012}
{Hlozek} R.,  et~al., 2012, \mn@doi [\apj] {10.1088/0004-637X/752/2/79}, \href {https://ui.adsabs.harvard.edu/abs/2012ApJ...752...79H} {752, 79}

\bibitem[\protect\citeauthoryear{Hochreiter \& Schmidhuber}{Hochreiter \& Schmidhuber}{1997}]{Hochreiter:1997}
Hochreiter S.,  Schmidhuber J.,  1997, \mn@doi [Neural Comput.] {10.1162/neco.1997.9.8.1735}, 9, 1735

\bibitem[\protect\citeauthoryear{{Jones} et~al.,}{{Jones} et~al.}{2018}]{Jones:2018}
{Jones} D.~O.,  et~al., 2018, \mn@doi [\apj] {10.3847/1538-4357/aab6b1}, \href {https://ui.adsabs.harvard.edu/abs/2018ApJ...857...51J} {857, 51}

\bibitem[\protect\citeauthoryear{{Kessler} et~al.,}{{Kessler} et~al.}{2009}]{Kessler:2009}
{Kessler} R.,  et~al., 2009, \mn@doi [\pasp] {10.1086/605984}, \href {https://ui.adsabs.harvard.edu/abs/2009PASP..121.1028K} {121, 1028}

\bibitem[\protect\citeauthoryear{{Kessler} et~al.,}{{Kessler} et~al.}{2010}]{Kessler:2010FittedPhotoZ}
{Kessler} R.,  et~al., 2010, \mn@doi [\apj] {10.1088/0004-637X/717/1/40}, \href {https://ui.adsabs.harvard.edu/abs/2010ApJ...717...40K} {717, 40}

\bibitem[\protect\citeauthoryear{{Kessler} et~al.,}{{Kessler} et~al.}{2015}]{Kessler:2015}
{Kessler} R.,  et~al., 2015, \mn@doi [\aj] {10.1088/0004-6256/150/6/172}, \href {https://ui.adsabs.harvard.edu/abs/2015AJ....150..172K} {150, 172}

\bibitem[\protect\citeauthoryear{{Kessler} et~al.,}{{Kessler} et~al.}{2019a}]{Kessler:2019simsplasticc}
{Kessler} R.,  et~al., 2019a, \mn@doi [\pasp] {10.1088/1538-3873/ab26f1}, \href {https://ui.adsabs.harvard.edu/abs/2019PASP..131i4501K} {131, 094501}

\bibitem[\protect\citeauthoryear{{Kessler} et~al.,}{{Kessler} et~al.}{2019b}]{Kessler:2019}
{Kessler} R.,  et~al., 2019b, \mn@doi [\mnras] {10.1093/mnras/stz463}, \href {https://ui.adsabs.harvard.edu/abs/2019MNRAS.485.1171K} {485, 1171}

\bibitem[\protect\citeauthoryear{{LSST Science Collaboration}}{{LSST Science Collaboration}}{2009}]{LSST:2009}
{LSST Science Collaboration} 2009, preprint, p. arXiv:0912.0201 (\mn@eprint {arXiv} {0912.0201})

\bibitem[\protect\citeauthoryear{Lasker}{Lasker}{2020}]{Lasker:thesis}
Lasker J.~E.,  2020, Phd thesis, University of Chicago, \url {https://doi.org/10.6082/uchicago.2744}

\bibitem[\protect\citeauthoryear{{Leoni}, {Ishida}, {Peloton}  \& {M{\"o}ller}}{{Leoni} et~al.}{2022}]{Leoni:2022}
{Leoni} M.,  {Ishida} E.~E.~O.,  {Peloton} J.,   {M{\"o}ller} A.,  2022, \mn@doi [\aap] {10.1051/0004-6361/202142715}, \href {https://ui.adsabs.harvard.edu/abs/2022A&A...663A..13L} {663, A13}

\bibitem[\protect\citeauthoryear{{Lidman} et~al.,}{{Lidman} et~al.}{2020}]{Lidman:2020}
{Lidman} C.,  et~al., 2020, \mn@doi [\mnras] {10.1093/mnras/staa1341}, \href {https://ui.adsabs.harvard.edu/abs/2020MNRAS.496...19L} {496, 19}

\bibitem[\protect\citeauthoryear{{Lochner}, {McEwen}, {Peiris}, {Lahav}  \& {Winter}}{{Lochner} et~al.}{2016}]{Lochner:2016}
{Lochner} M.,  {McEwen} J.~D.,  {Peiris} H.~V.,  {Lahav} O.,   {Winter} M.~K.,  2016, \mn@doi [\apjs] {10.3847/0067-0049/225/2/31}, \href {https://ui.adsabs.harvard.edu/abs/2016ApJS..225...31L} {225, 31}

\bibitem[\protect\citeauthoryear{{Mitra}, {Kessler}, {More}, {Hlozek}  \& {LSST Dark Energy Science Collaboration}}{{Mitra} et~al.}{2023}]{Mitra:2023}
{Mitra} A.,  {Kessler} R.,  {More} S.,  {Hlozek} R.,   {LSST Dark Energy Science Collaboration} 2023, \mn@doi [\apj] {10.3847/1538-4357/acb057}, \href {https://ui.adsabs.harvard.edu/abs/2023ApJ...944..212M} {944, 212}

\bibitem[\protect\citeauthoryear{M{\"o}ller \& de Boissi{\`e}re}{M{\"o}ller \& de~Boissi{\`e}re}{2019}]{Moller:2020}
M{\"o}ller A.,  de Boissi{\`e}re T.,  2019, \mn@doi [MNRAS] {10.1093/mnras/stz3312}, \href {https://ui.adsabs.harvard.edu/abs/2019arXiv190106384M} {491, 4277}

\bibitem[\protect\citeauthoryear{{M{\"o}ller} et~al.,}{{M{\"o}ller} et~al.}{2016}]{Moller:2016}
{M{\"o}ller} A.,  et~al., 2016, \mn@doi [Journal of Cosmology and Astro-Particle Physics] {10.1088/1475-7516/2016/12/008}, \href {https://ui.adsabs.harvard.edu/#abs/2016JCAP...12..008M} {2016, 008}

\bibitem[\protect\citeauthoryear{{M{\"o}ller} et~al.,}{{M{\"o}ller} et~al.}{2021}]{Moller:2021}
{M{\"o}ller} A.,  et~al., 2021, \mn@doi [\mnras] {10.1093/mnras/staa3602}, \href {https://ui.adsabs.harvard.edu/abs/2021MNRAS.501.3272M} {501, 3272}

\bibitem[\protect\citeauthoryear{{Muthukrishna}, {Narayan}, {Mandel}, {Biswas}  \& {Hlo{\v{z}}ek}}{{Muthukrishna} et~al.}{2019}]{Muthukrishna:2019}
{Muthukrishna} D.,  {Narayan} G.,  {Mandel} K.~S.,  {Biswas} R.,   {Hlo{\v{z}}ek} R.,  2019, \mn@doi [\pasp] {10.1088/1538-3873/ab1609}, \href {https://ui.adsabs.harvard.edu/abs/2019PASP..131k8002M} {131, 118002}

\bibitem[\protect\citeauthoryear{{Möller} \& {Main de Boissière}}{{Möller} \& {Main de Boissière}}{2022}]{Moller:2022BNN}
{Möller} A.,  {Main de Boissière} T.,  2022, in Machine Learning for Astrophysics. p.~21 (\mn@eprint {arXiv} {2207.04578}), \mn@doi{10.48550/arXiv.2207.04578}

\bibitem[\protect\citeauthoryear{Möller et~al.,}{Möller et~al.}{2022}]{Moller:2022}
Möller A.,  et~al., 2022, \mn@doi [MNRAS] {10.1093/mnras/stac1691}

\bibitem[\protect\citeauthoryear{{Palanque-Delabrouille} et~al.,}{{Palanque-Delabrouille} et~al.}{2010}]{Palanque-Delabrouille:2010}
{Palanque-Delabrouille} N.,  et~al., 2010, \mn@doi [\aap] {10.1051/0004-6361/200913283}, \href {https://ui.adsabs.harvard.edu/abs/2010A&A...514A..63P} {514, A63}

\bibitem[\protect\citeauthoryear{{Pierel} et~al.,}{{Pierel} et~al.}{2018}]{Pierel:2018}
{Pierel} J.~D.~R.,  et~al., 2018, \mn@doi [\pasp] {10.1088/1538-3873/aadb7a}, \href {https://ui.adsabs.harvard.edu/abs/2018PASP..130k4504P} {130, 114504}

\bibitem[\protect\citeauthoryear{{Planck Collaboration}}{{Planck Collaboration}}{2020}]{Planck:2018}
{Planck Collaboration} 2020, \mn@doi [\aap] {10.1051/0004-6361/201833910}, \href {https://ui.adsabs.harvard.edu/abs/2020A&A...641A...6P} {641, A6}

\bibitem[\protect\citeauthoryear{{Qu} \& {Sako}}{{Qu} \& {Sako}}{2023}]{Qu:2023}
{Qu} H.,  {Sako} M.,  2023, \mn@doi [\apj] {10.3847/1538-4357/aceafa}, \href {https://ui.adsabs.harvard.edu/abs/2023ApJ...954..201Q} {954, 201}

\bibitem[\protect\citeauthoryear{{Qu}, {Sako}, {M{\"o}ller}  \& {Doux}}{{Qu} et~al.}{2021}]{Qu:2021}
{Qu} H.,  {Sako} M.,  {M{\"o}ller} A.,   {Doux} C.,  2021, \mn@doi [\aj] {10.3847/1538-3881/ac0824}, \href {https://ui.adsabs.harvard.edu/abs/2021AJ....162...67Q} {162, 67}

\bibitem[\protect\citeauthoryear{{Ruhlmann-Kleider}, {Lidman}  \& {M{\"o}ller}}{{Ruhlmann-Kleider} et~al.}{2022}]{Ruhlmann-Kleider:2022}
{Ruhlmann-Kleider} V.,  {Lidman} C.,   {M{\"o}ller} A.,  2022, \mn@doi [\jcap] {10.1088/1475-7516/2022/10/065}, \href {https://ui.adsabs.harvard.edu/abs/2022JCAP...10..065R} {2022, 065}

\bibitem[\protect\citeauthoryear{Sako et~al.,}{Sako et~al.}{2011}]{Sako:2011}
Sako M.,  et~al., 2011, \mn@doi [The Astrophysical Journal] {10.1088/0004-637x/738/2/162}, 738, 162

\bibitem[\protect\citeauthoryear{{Scolnic} et~al.,}{{Scolnic} et~al.}{2018}]{Scolnic:2018}
{Scolnic} D.~M.,  et~al., 2018, \mn@doi [\apj] {10.3847/1538-4357/aab9bb}, \href {https://ui.adsabs.harvard.edu/abs/2018ApJ...859..101S} {859, 101}

\bibitem[\protect\citeauthoryear{{Smith} et~al.,}{{Smith} et~al.}{2018}]{Smith:2018}
{Smith} M.,  et~al., 2018, \mn@doi [\apj] {10.3847/1538-4357/aaa126}, \href {https://ui.adsabs.harvard.edu/abs/2018ApJ...854...37S} {854, 37}

\bibitem[\protect\citeauthoryear{Smith et~al.,}{Smith et~al.}{2020}]{Smith:2020}
Smith M.,  et~al., 2020, \mn@doi [The Astronomical Journal] {10.3847/1538-3881/abc01b}, 160, 267

\bibitem[\protect\citeauthoryear{{Swann} et~al.,}{{Swann} et~al.}{2019}]{Swann:2019}
{Swann} E.,  et~al., 2019, \mn@doi [The Messenger] {10.18727/0722-6691/5129}, \href {https://ui.adsabs.harvard.edu/abs/2019Msngr.175...58S} {175, 58}

\bibitem[\protect\citeauthoryear{{Villar} et~al.,}{{Villar} et~al.}{2019}]{Villar:2019}
{Villar} V.~A.,  et~al., 2019, \mn@doi [\apj] {10.3847/1538-4357/ab418c}, \href {https://ui.adsabs.harvard.edu/abs/2019ApJ...884...83V} {884, 83}

\bibitem[\protect\citeauthoryear{{Villar} et~al.,}{{Villar} et~al.}{2020}]{Villar:2020}
{Villar} V.~A.,  et~al., 2020, \mn@doi [\apj] {10.3847/1538-4357/abc6fd}, \href {https://ui.adsabs.harvard.edu/abs/2020ApJ...905...94V} {905, 94}

\bibitem[\protect\citeauthoryear{{Vincenzi}, {Sullivan}, {Firth}, {Guti{\'e}rrez}, {Frohmaier}, {Smith}, {Angus}  \& {Nichol}}{{Vincenzi} et~al.}{2019}]{Vincenzi:2019}
{Vincenzi} M.,  {Sullivan} M.,  {Firth} R.~E.,  {Guti{\'e}rrez} C.~P.,  {Frohmaier} C.,  {Smith} M.,  {Angus} C.,   {Nichol} R.~C.,  2019, \mn@doi [\mnras] {10.1093/mnras/stz2448}, \href {https://ui.adsabs.harvard.edu/abs/2019MNRAS.489.5802V} {489, 5802}

\bibitem[\protect\citeauthoryear{{Vincenzi} et~al.,}{{Vincenzi} et~al.}{2021}]{Vincenzi:2020}
{Vincenzi} M.,  et~al., 2021, \mn@doi [\mnras] {10.1093/mnras/stab1353}, \href {https://ui.adsabs.harvard.edu/abs/2021MNRAS.505.2819V} {505, 2819}

\bibitem[\protect\citeauthoryear{{Vincenzi} et~al.,}{{Vincenzi} et~al.}{2022}]{Vincenzi:2022}
{Vincenzi} M.,  et~al., 2022, \mn@doi [\mnras] {10.1093/mnras/stac1404}, \href {https://ui.adsabs.harvard.edu/abs/2022MNRAS.tmp.1479V} {}

\bibitem[\protect\citeauthoryear{{Vincenzi} et~al.,}{{Vincenzi} et~al.}{2024}]{Vincenzi:2024}
{Vincenzi} M.,  et~al., 2024, \mn@doi [arXiv e-prints] {10.48550/arXiv.2401.02945}, \href {https://ui.adsabs.harvard.edu/abs/2024arXiv240102945V} {p. arXiv:2401.02945}

\bibitem[\protect\citeauthoryear{{Wiseman} et~al.,}{{Wiseman} et~al.}{2020a}]{Wiseman:2020}
{Wiseman} P.,  et~al., 2020a, \mn@doi [\mnras] {10.1093/mnras/staa1302}, \href {https://ui.adsabs.harvard.edu/abs/2020MNRAS.495.4040W} {495, 4040}

\bibitem[\protect\citeauthoryear{{Wiseman} et~al.,}{{Wiseman} et~al.}{2020b}]{Wiseman:2020a}
{Wiseman} P.,  et~al., 2020b, \mn@doi [\mnras] {10.1093/mnras/staa2474}, \href {https://ui.adsabs.harvard.edu/abs/2020MNRAS.498.2575W} {498, 2575}

\bibitem[\protect\citeauthoryear{{Yuan} et~al.,}{{Yuan} et~al.}{2015}]{Yuan:2015}
{Yuan} F.,  et~al., 2015, \mn@doi [\mnras] {10.1093/mnras/stv1507}, \href {https://ui.adsabs.harvard.edu/abs/2015MNRAS.452.3047Y} {452, 3047}

\makeatother
\end{thebibliography}



 \appendix

\section{DES HQ sample and SNphoto-z}\label{appendix:sample_biases_photoz}
In this appendix, we inspect events in the HQ sample introduced in Section~\ref{sec:HQsample} and their SNphoto-z and SALT2 light-curve parameter fits.

\begin{figure*}
    \centering
    \includegraphics[width=\textwidth]{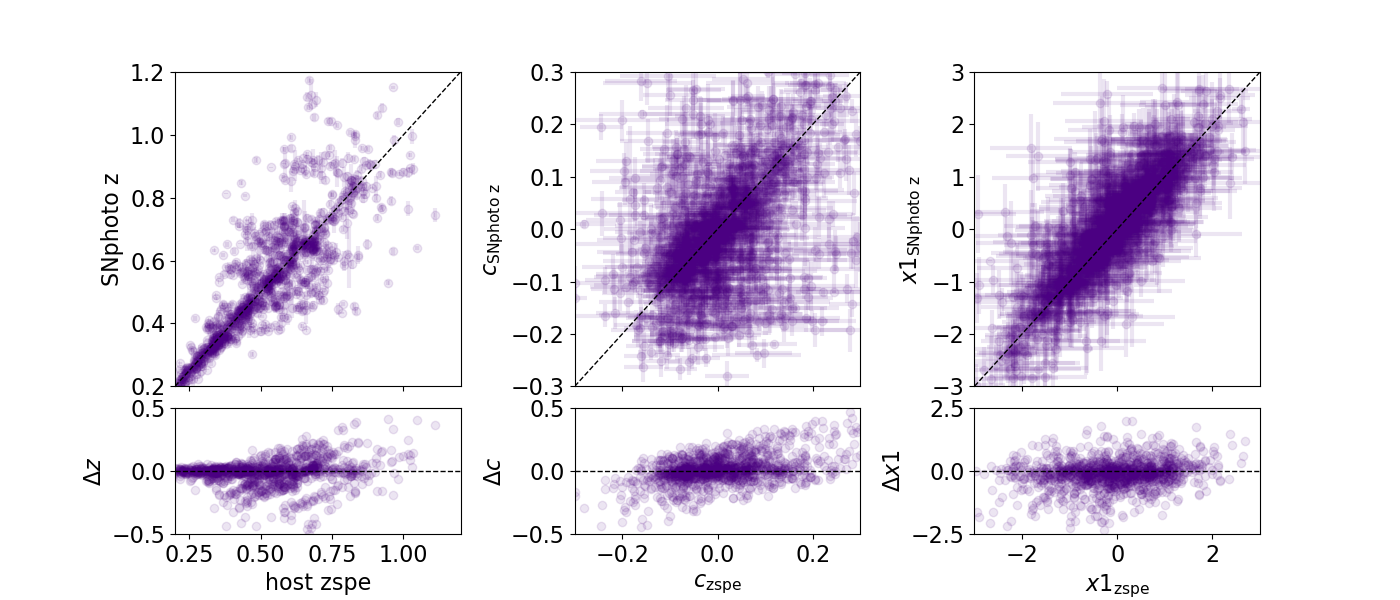}
    \caption{DES HQ overlap with \citetalias{Moller:2022}. Left: a comparison of the SNphoto-z versus host-galaxy spectroscopic redshifts. Center and right: comparisons between light-curve parameters colour ($c$) and stretch ($x1$) using the SNphoto-zs and the host spectroscopic one. The dashed line shows the diagonal where the values should converge if they were equivalent. The lower row indicated the difference between parameters with host-galaxy spectroscopic redshift and SNphoto-z. Events classified by SNe Ia by both methods are mostly consistent with their estimates.}
    \label{fig:delta_retro_overlap}
\end{figure*}

\begin{figure*}
    \centering
    \includegraphics[width=\textwidth]{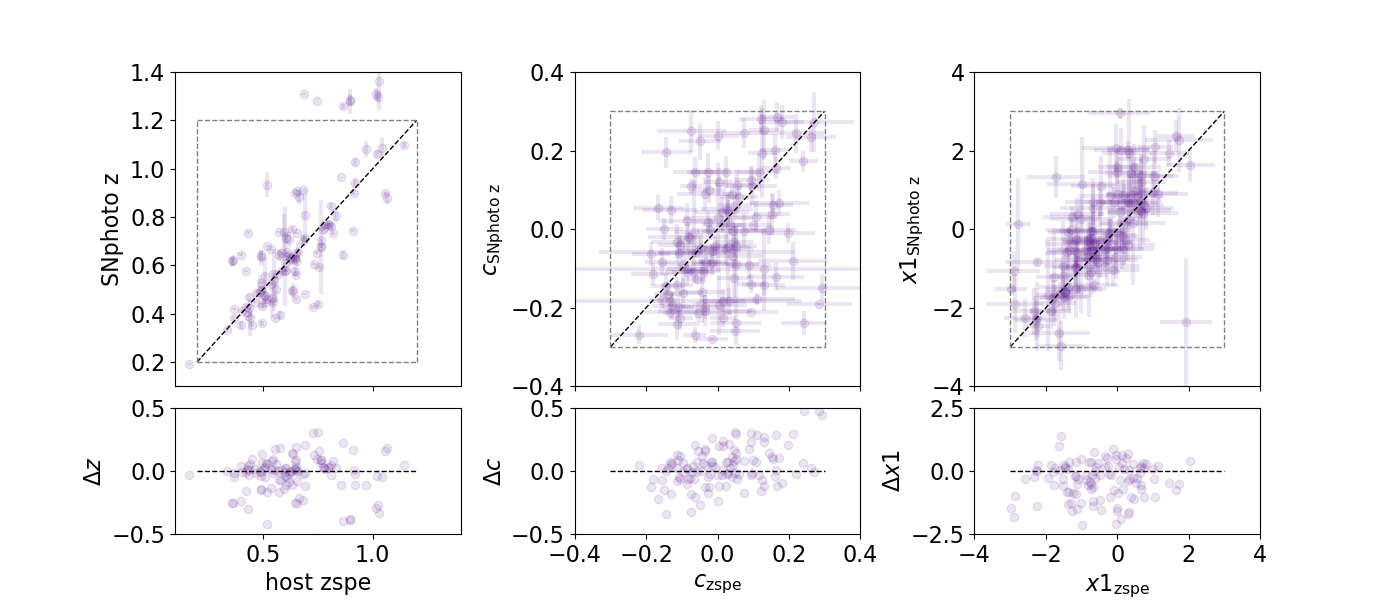}
    \caption{\citetalias{Moller:2022} SNe Ia not included in the DES HQ sample (248 SNe Ia). We show only events that have a converging SNphoto z fit (116 events). Left: a comparison of the SNphoto-z versus host-galaxy spectroscopic redshifts. Center and right: comparisons between light-curve parameters colour ($c$) and stretch ($x1$) using the SNphoto-zs and the host spectroscopic one. The dashed line shows the diagonal where the values should converge if they were equivalent. Lost \citetalias{Moller:2022} events include SNe Ia which have SNphoto-z beyond the HQ cuts (shown as a square grey dashed line) and some more scattered fitted events.}
    \label{fig:delta_retro_lost}
\end{figure*}

In Figure~\ref{fig:delta_retro_overlap} we find that for the common events in DES HQ SNe Ia and the \citetalias{Moller:2022} sample, the SNphoto-z estimation agrees mostly with their spectroscopic host redshifts. For the 248 events from \citetalias{Moller:2022} that are not selected in our z-free sample, only 116 obtain a SNphoto z estimation. For the latter, in Figure~\ref{fig:delta_retro_lost} we find a large dispersion on the fitted vs. spectroscopic redshift parameters. In some cases these parameters are estimated outside the HQ cuts.

\section{Fitted light-curve photometric redshift}\label{appendix:distancemusalt}

The method used in this work to estimate photometric redshifts by simultaneously fitting redshift with SALT2 light-curve parameters is further described in \cite{Kessler:2010FittedPhotoZ} and in the SNANA manual Section 5.12. 

In this appendix, we clarify the distance prior mechanism used for this fit. We assume a $\Lambda CDM$ cosmology with wide priors centred around $H_0 = 70$, $w=-1$, $\Omega_{m} = 0.315$, $\Omega_\Lambda =0.685$ \citep{Planck:2018}.

First, the fitted distance modulus, $\mu_{SALT2}$, is approximately computed as
   \begin{equation}
   \mu_{SALT2} = 30.0 - 2.5*log10(x0) + (\alpha *x1) - (\beta * c)
   \end{equation}
where $x0, x1$ and $c$ are SALT2 light-curve parameters and we use default parameters $\alpha = 0.14$ and $\beta  = 3.2$.

Next, the difference between the fitted and theoretical distance modulus, $\mu_{DIF}$, is computed as:
\begin{equation}
   \mu_{DIF} = \mu_{SALT2} -  \mu_{\rm theory}(z_{PHOT}) 
\end{equation}
where $z_{PHOT}$ is the SNphoto-z.

An intentionally large estimate of the distance uncertainty, $\sigma_{\mu}^2$, is given by:
\begin{equation}
   \sigma_{\mu}^2 = 4([d\mu/d\Omega_M * \sigma_{\Omega_M}]^2 + [d\mu/dw * \sigma_w]^2)
\end{equation}
where $\sigma_{\Omega_M}=0.3$ and $\sigma_w=0.1$ are errors in the cosmological parameters, the factor 4 is an arbitrary number to do an overestimation of the uncertainty, $\Omega_M$ and $w$, matter density and equation of state of Dark Energy respectively. 

For the fitting procedure, to the nominal SALT2 $\chi^2$, we add: 
\begin{equation}
   \Delta \chi^2_{SALT2} = [\mu_{DIF} / \sigma_\mu ]^2
\end{equation}

\section{Author Affiliations}\label{appendix:affiliations}
$^{1}$ Centre for Astrophysics \& Supercomputing, Swinburne University of Technology, Victoria 3122, Australia \\ 
$^{2}$ ARC Centre of Excellence for Gravitational Wave Discovery (OzGrav), Australia \\ 
$^{3}$ School of Physics and Astronomy, University of Southampton,  Southampton, SO17 1BJ, UK \\ 
$^{4}$ Physics Department, Lancaster University, Lancaster, LA1 4YB, UK \\ 
$^{5}$ The Research School of Astronomy and Astrophysics, Australian National University, ACT 2601, Australia \\ 
$^{6}$ Centre for Gravitational Astrophysics, College of Science, The Australian National University, ACT 2601, Australia \\ 
$^{7}$ School of Mathematics and Physics, University of Queensland,  Brisbane, QLD 4072, Australia \\ 
$^{8}$ Kavli Institute for Cosmological Physics, University of Chicago, Chicago, IL 60637, USA \\ 
$^{9}$ Department of Astronomy and Astrophysics, University of Chicago, Chicago, IL 60637, USA \\ 
$^{10}$ Department of Physics and Astronomy, University of Pennsylvania, Philadelphia, PA 19104, USA \\ 
$^{11}$ Institute of Space Sciences (ICE, CSIC),  Campus UAB, Carrer de Can Magrans, s/n,  08193 Barcelona, Spain \\ 
$^{12}$ Institut d'Estudis Espacials de Catalunya (IEEC), 08034 Barcelona, Spain \\ 
$^{13}$ School of Mathematics and Physics, University of Surrey, Guildford, UK \\ 
$^{14}$ Aix Marseille Univ, CNRS/IN2P3, CPPM, Marseille, France \\ 
$^{15}$ Department of Physics, Duke University Durham, NC 27708, USA \\ 
$^{16}$ Cerro Tololo Inter-American Observatory, NSF's National Optical-Infrared Astronomy Research Laboratory, Casilla 603, La Serena, Chile \\ 
$^{17}$ Laborat\'orio Interinstitucional de e-Astronomia - LIneA, Rua Gal. Jos\'e Cristino 77, Rio de Janeiro, RJ - 20921-400, Brazil \\ 
$^{18}$ Fermi National Accelerator Laboratory, P. O. Box 500, Batavia, IL 60510, USA \\ 
$^{19}$ Department of Physics, University of Michigan, Ann Arbor, MI 48109, USA \\ 
$^{20}$ Institute of Cosmology and Gravitation, University of Portsmouth, Portsmouth, PO1 3FX, UK \\ 
$^{21}$ CNRS, UMR 7095, Institut d'Astrophysique de Paris, F-75014, Paris, France \\ 
$^{22}$ Sorbonne Universit\'es, UPMC Univ Paris 06, UMR 7095, Institut d'Astrophysique de Paris, F-75014, Paris, France \\ 
$^{23}$ Department of Physics \& Astronomy, University College London, Gower Street, London, WC1E 6BT, UK \\ 
$^{24}$ Universidad de La Laguna, Dpto. Astrofísica, E-38206 La Laguna, Tenerife, Spain \\ 
$^{25}$ Instituto de Astrofisica de Canarias, E-38205 La Laguna, Tenerife, Spain \\ 
$^{26}$ Department of Physics, IIT Hyderabad, Kandi, Telangana 502285, India \\ 
$^{27}$ Jet Propulsion Laboratory, California Institute of Technology, 4800 Oak Grove Dr., Pasadena, CA 91109, USA \\ 
$^{28}$ Institute of Theoretical Astrophysics, University of Oslo. P.O. Box 1029 Blindern, NO-0315 Oslo, Norway \\ 
$^{29}$ Center for Astrophysical Surveys, National Center for Supercomputing Applications, 1205 West Clark St., Urbana, IL 61801, USA \\ 
$^{30}$ Instituto de Fisica Teorica UAM/CSIC, Universidad Autonoma de Madrid, 28049 Madrid, Spain \\ 
$^{31}$ Institut de F\'{\i}sica d'Altes Energies (IFAE), The Barcelona Institute of Science and Technology, Campus UAB, 08193 Bellaterra (Barcelona) Spain \\ 
$^{32}$ Department of Astronomy, University of Illinois at Urbana-Champaign, 1002 W. Green Street, Urbana, IL 61801, USA \\ 
$^{33}$ Santa Cruz Institute for Particle Physics, Santa Cruz, CA 95064, USA \\ 
$^{34}$ Department of Physics, The Ohio State University, Columbus, OH 43210, USA \\ 
$^{35}$ Center for Cosmology and Astro-Particle Physics, The Ohio State University, Columbus, OH 43210, USA \\ 
$^{36}$ Center for Astrophysics $\vert$ Harvard \& Smithsonian, 60 Garden Street, Cambridge, MA 02138, USA \\ 
$^{37}$ Lowell Observatory, 1400 Mars Hill Rd, Flagstaff, AZ 86001, USA \\ 
$^{38}$ Australian Astronomical Optics, Macquarie University, North Ryde, NSW 2113, Australia \\ 
$^{39}$ George P. and Cynthia Woods Mitchell Institute for Fundamental Physics and Astronomy, and Department of Physics and Astronomy, Texas A\&M University, College Station, TX 77843,  USA \\ 
$^{40}$ LPSC Grenoble - 53, Avenue des Martyrs 38026 Grenoble, France \\ 
$^{41}$ Instituci\'o Catalana de Recerca i Estudis Avan\c{c}ats, E-08010 Barcelona, Spain \\ 
$^{42}$ Department of Astrophysical Sciences, Princeton University, Peyton Hall, Princeton, NJ 08544, USA \\ 
$^{43}$ Observat\'orio Nacional, Rua Gal. Jos\'e Cristino 77, Rio de Janeiro, RJ - 20921-400, Brazil \\ 
$^{44}$ Department of Physics, Carnegie Mellon University, Pittsburgh, Pennsylvania 15312, USA \\ 
$^{45}$ SLAC National Accelerator Laboratory, Menlo Park, CA 94025, USA \\ 
$^{46}$ Kavli Institute for Particle Astrophysics \& Cosmology, P. O. Box 2450, Stanford University, Stanford, CA 94305, USA \\ 
$^{47}$ Centro de Investigaciones Energ\'eticas, Medioambientales y Tecnol\'ogicas (CIEMAT), Madrid, Spain \\ 
$^{48}$ Computer Science and Mathematics Division, Oak Ridge National Laboratory, Oak Ridge, TN 37831 \\ 
$^{49}$ Lawrence Berkeley National Laboratory, 1 Cyclotron Road, Berkeley, CA 94720, USA \\ 
$^{50}$ Hamburger Sternwarte, Universit\"{a}t Hamburg, Gojenbergsweg 112, 21029 Hamburg, Germany


\bsp    
\label{lastpage}
\end{document}